\documentclass[12pt]{article}
\usepackage{kpfonts}
 \usepackage{graphicx}  
\usepackage{amsmath}   
\usepackage{bm}
\usepackage[compress]{cite}
\usepackage{amssymb}   
\usepackage{bm} 
\usepackage{dcolumn}
\usepackage{ulem}
\usepackage{color}
\usepackage{mathtools}
\usepackage{mathrsfs}
\usepackage{amsfonts}
\usepackage{varioref}
\usepackage{textcomp}
\usepackage{tcolorbox}
\usepackage{empheq}
\usepackage[active]{srcltx}
\usepackage{tikz}
\usepackage{lmodern}
\usepackage{braket}
\usetikzlibrary{decorations.pathmorphing}
\tikzset{snake it/.style={decorate, decoration=snake}}
\usepackage{fancybox}
\usepackage{tikz,caption}
\usetikzlibrary{shapes.geometric, calc,positioning}
\usetikzlibrary{patterns}
\usetikzlibrary{backgrounds}
\usepackage{placeins}
\usepackage{dirtytalk}
\date{} 
\RequirePackage[colorlinks,citecolor=blue,urlcolor=blue,linkcolor=blue]{hyperref}
\allowdisplaybreaks
\usepackage{hyperref}
\hypersetup{colorlinks,allcolors=blue}
\usepackage[top=80pt,bottom=85pt,left=60pt,right=60pt]{geometry}
\usepackage{overpic}

\def\be{\begin{equation}}
\def\bea{\begin{eqnarray}}
\def\ee{\end{equation}}
\def\eea{\end{eqnarray}}

\numberwithin{equation}{section}

\definecolor{newblue}{rgb}{0,0,.6}
\definecolor{newred}{RGB}{178,34,34}
\definecolor{newgreen}{rgb}{0.18, 0.65, 0.3}
\definecolor{newgray}{RGB}{101,101,101}
\definecolor{newyellow}{rgb}{.96,.78,.10}
\definecolor{newpurple}{RGB}{128,0,128}
\definecolor{newteal}{RGB}{0,128,128}
\definecolor{newpink}{rgb}{1,.49,.47}

\usepackage{tikz}
\usetikzlibrary{shapes.misc}

\usepackage{tikz-3dplot}
\definecolor{mycolor1}{rgb}{0.9, 0.1, 0.4}

\begin{document}

\begin{titlepage}

\hspace*{\fill} CQUeST-2026-07779

\begin{center}

\vskip .3in 
\noindent

{\Large \bf{BCFW like recursion for Deformed Associahedron}}

\bigskip\medskip

Sujoy Mahato${}^{a^\star}$  and Sourav Roychowdhury${}^{b^\dagger}$  \\

\bigskip\medskip
{\small 
 
 ${}^a${\it{Harish-Chandra Research Institute\\
Chhatnag Road, Jhunsi, Allahabad 211019, India}}\\[2mm]


 \vskip .2in
 
 ${}^b${\it{Department of Physics \& Center for Quantum Spacetime, Sogang University\\
35 Baekbeom-ro, Mapo-gu, Seoul 04107, Republic of Korea}}\\[2mm]

}

\bigskip\medskip

\vskip 0cm 
{\small \tt ${}^\star$sujoymahato@hri.res.in, ${}^\dagger$srcphys@sogang.ac.kr}
\vskip .9cm 
     	{\bf Abstract }

\vskip .1in
\end{center}
\noindent

In this paper, we explore the applicability of the BCFW-like recursion relations \cite{He:2018svj,Yang:2019esm} to a  wider class of positive geometries. Previously it was found in \cite{Jagadale:2022rbl}, the tree level scattering amplitude of a theory with more than one type of scalar particles interacting via cubic couplings of different strength can be captured by a deformed realization of the ABHY-associahedron in the kinematic space. In the literature, we explore the adaptation of the recursion relations for the case of deformed associahedron. The formalism is further generalized to the deformed realization of the D-type cluster polytopes which captures the one-loop amplitudes in this class of cubic theories. These recursion terms correspond to projective triangulation of the associahedron (or D-type cluster polytopes). Towards the end, we briefly mention the idea of recovering EFT amplitudes from the cubic theory in terms of recursion relations.

\vfill
\eject

\end{titlepage}

\tableofcontents

\section{Introduction and summary } \label{int}

Scattering amplitude and its relation
 to mathematical and geometric structures is an important arena of research for the last couple of decades. 
The main theme of this framework is to understand the corresponding geometric structure as an abstract object from which one can derive the scattering amplitudes. The main upshot of this program is a radical change in the concept of locality and unitarity in quantum field theories, rather than taking them as the guiding principles to write down a spacetime local Lagrangian, these sacred properties emerge out of the boundary structure of these abstract geometric objects. The first of this kind of geometric structure was found in the context of scattering amplitude of a special kind of theory namely the  $\mathcal N=4$ SYM theory and was termed as the {\it Amplitudehron} \cite{Arkani-Hamed:2013jha}. It lives in the abstract space of positive  Grassmannian. 
There was an uneasy gap in literature to relate these geometric structures in auxiliary space to S-matrix in kinematic space.
The puzzle was first resolved by Arkami-Hamed, Bai, He and Yan (ABHY) in \cite{Arkani-Hamed:2017mur} for the case in tree-level amplitudes in bi-adjoint $\phi^3$ theory by introducing canonical forms on a special class of  positive geometry namely  {\it{associahedron}} that lives on the kinematic space\footnote{Recently, an alternative positive geometry called the {\it Momentum Amplituhedron} was proposed to capture the scattering amplitudes of $\mathcal N=4$ SYM theory and it lives in the spinor-helicity space \cite{Damgaard:2019ztj}. Further, a connection between the momentum amplituhedron and kinematic associahedron was explored in \cite{Damgaard:2020eox}.}. 
Later on, this program also revealed for planar tree-level amplitudes in $\phi^4$ theory \cite{Banerjee:2018tun}  as well as $\phi^p$ theories with $p>4$ \cite{Raman:2019utu} and the polynomial interactions such as $\phi^3 + \phi^4$ \cite{Aneesh:2019ddi}. For the former case the positive geometry is known as {\it{Stokes polytope}} whereas for the later two cases it is a polytope known as accordiohedron. Amplitudes with matter particles was included in the positive geometry program in \cite{Herderschee:2019wtl}; the corresponding polytope is unbounded and known as {\it open associahedra}. Since it's conception, this program was also extended for loop-level amplitudes in scalar theories over the years \cite{Arkani-Hamed:2019vag,Salvatori:2018aha,Jagadale:2020qfa}. For the above developments, connections to graph theory and cluster algebras have proven to be useful \cite{Bazier-Matte:2018rat,Padrol:2019zsr,Ceballos_2015,Kalyanapuram:2020vil}.
Very recently attempts were made to extend it to gauge theories \cite{Laddha:2024qtn} and theories with gravity \cite{Trnka:2020dxl}. The object that acts as a positive geometry for cosmological correlators is the {\it Cosmological polytope}\cite{Arkani-Hamed:2017fdk,Benincasa:2019vqr} and {\it Cosmohedra} \cite{Arkani-Hamed:2024jbp}.  A very interesting extension towards the string amplitudes via. various stringy canonical forms could be found in \cite{Arkani-Hamed:2019mrd,He:2020onr,Cao:2025lzv,Huang:2020nqy,He:2020ray}. This rapid growing pool of literature points towards both the conceptual importance as well as the wide applicability of the idea of geometrization of scattering amplitudes for a wide class of QFTs (with or without supersymmetry) and beyond \cite{Arkani-Hamed:2018ign,Arkani-Hamed:2020blm,Figueiredo:2025daa,Glew:2025otn,Forcey:2025voc,He:2021llb,Lukowski:2023nnf}.

Recently the program was extended to a most generalized class of theories containing more than one type of scalar particles interacting with each other through cubic couplings of different strengths \cite{Jagadale:2022rbl}. The corresponding positive geometry turned out to be a deformed realization of the associahedron in the kinematic space through a linear deformation of the kinematic planar variables with different parameters. These deformation parameters are eventually related to the various coupling strengths of the three point interactions.

The structure of S-matrix is special in the sense it is an on-shell object and in principal can be derived with the data of in-states and out-states.  In many cases one can bypass the calculational complications  arising from Feynman diagrams which critically depend upon off-shell quantities like Lagrangian. Recent developments such as BCFW on-shell recursion relations \cite{Britto:2005fq,Britto:2004ap,Arkani-Hamed:2008bsc} has successfully used the analytical structure of the S-matrix to great effect in computing the scattering amplitude with higher efficiency. Alternatively, one can derive the scattering amplitude for a class of theories from a specific compactification of the worldsheet moduli space viz. the famous CHY formula \cite{Cachazo:2013iea}. The natural curiosity arise about these two seemingly different approaches to calculate the amplitude is related to the geometric approach of polytopes. This was partially addressed in \cite{Arkani-Hamed:2017mur} that the canonical top form on an {\it worldsheet associahedron} located inside the moduli space and the top form onto the {\it kinematic associahedron} are related by a formula equivalent to CHY \footnote{Similar CHY-like representation was found for quartic theories in \cite{Aneesh:2019cvt}.}. From a geometric viewpoint, these findings have shed new light on the worldsheet approach to scattering amplitudes. 

In the same ABHY paper, they also commented upon a recursive scheme to obtain the canonical form for a higher dimensional associahedron by gluing the lower dimensional ones. It was considered an equivalent study of the BCFW relation for the case of positive geometries through a series of works starting from $\phi^3$ theory \cite{He:2018svj,Yang:2019esm} and then extended for $\phi^p$ theories in \cite{Salvatori:2019phs,Kojima:2020tox,John:2020jww,Kalyanapuram:2020tsr}.
The outline of this framework is as follows. 
First, we introduce a complex variable $(z)$ and scale all basis planar variables $X_{ij}$ as $X_{ij} \rightarrow zX_{ij}$ along with constructing a meromorphic function that does not have any pole at $z=0, \infty$. 
This formalism precedes a similar construction in supersymmetric amplitudes  \cite{Brandhuber:2008pf,Arkani-Hamed:2008owk,Elvang:2008na,Drummond:2008cr,Bianchi:2008pu,Cheung:2008dn,Arkani-Hamed:2010zjl,Huang:2021jlh}.
An analogous recursion framework was put forward for open associahedra in \cite{Herderschee:2020lgb} making the recursion relations applicable for a wide variety of positive geometries.

In the recursion, each term corresponds to the canonical function of one of the sub-geometries that obtained by the full triangulations of the associahedron, in similar spirit to the BCFW representation of the $\mathcal N=4$ SYM theory \cite{Hodges:2009hk,Arkani-Hamed:2010wgm}. The upshot of this whole discussion is that the approach of positive geometries to scattering amplitude gives a geometric interpretation to the different concepts or approaches leading to the same quantity.

In the case of cubic theory, the recursion relations project the associahedron onto one of its facets, known as projective triangulation. 
In some special cases, projective triangulation emerges the interesting property that it divides the parent geometry into positive geometries containing curvy facts.
The recursion relation that corresponds to such triangulations were derived from the field theory in the case of tree as well as one loop amplitudes of the bi-adjoint $\phi^3$ theory. 
In this paper we investigated this projective triangulation for the case of deformed planar as well as deformed loop amplitude.

The paper is organized as follows.
To begin with, in section\eqref{review}, we briefly review the basic ingredients about the construction of the deformed associahedron that are needed for our work.

In section\eqref{recursion}, we discuss the recursion relation for deformed associahedron and work out some specific cases of deformation.  
In particular, we discuss the deformed four-point and six-point amplitude in massless $\phi^3$ theory.
In the case of six-point amplitude we rescaled two basis planar variables and along the same line, we also discussed the recursion relation in one-loop amplitude. 
Afterwards in section\eqref{intep}, we provide the interpretation of our findings in \eqref{recursion} in a more general framework.
In section\eqref{geo} we workout the geometric pictures of the recursion terms in tree level and loop case through considering projective triangulations of the deformed associahedron. 

Towards the end, in section\eqref{EFT} we discuss the effective field theory (EFT) that emerges from the various limits of the coupling constants appearing in the interaction terms of the Lagrangian. This fact leads to particular kind of limits for the recursion terms  of the deformed associahedron which connects to the EFT amplitude.

Finally in section\eqref{con} we conclude and provide some future avenues to pursue later. 
We provide some important supplementary materials in the appendices.

\section{Review of the associahedron as the amplituhedron of general scalar theory} \label{review}

To begin with, we briefly review the construction of the 
amplituhedron in the case of planar scalar field theories based on \cite{Arkani-Hamed:2017mur,Banerjee:2018tun,Raman:2019utu,Aneesh:2019ddi,Aneesh:2019cvt,Jagadale:2021iab,Yang:2019esm,Herrmann:2022nkh}. 
 Focusing the tree level scattering amplitude in a planar theory, schematically the pole structures (in other words the propagators) are given by 
 $\frac{1}{X_{ij}-m_{ij}^2}$, where $X_{ij} = (p_{i}+ p_{i+1} + \cdots + p_{j-1} )^{2}$ and $p_{i}$ is momenta of the $i$-th external particle.
 But it turns out all poles do not appear in a single term in the amplitude.
 For example, in the case of 5-point amplitude, we do not find the poles $\frac{1}{X_{13}-m_{13}^2}$ and $\frac{1}{X_{24}-m_{24}^2}$ together in the same Feynman diagram. 
 The above property implies that the set of all possible poles in the planar tree level $n$-point scattering amplitude containing with one scalar field, has a structure of a well known combinatorial polytope namely the associahedron.

Interestingly, the associahedron plays a role not only for a theory with cubic interaction but also with theories having a more general kind of interaction. Consider the pole $\frac{1}{X_{13}-m_{13}^2}$, which can appear for a $\phi^3$ theory but never in a $\phi^{p \ge 4}$ theory. In short, in the theories with cubic interaction exhaust all the pole structure that is possible for $n$-point amplitude of a tree level planar theory. Hence, the associahedron being the positive geometry of a theory that contains a cubic interaction plays a central role in the formulation for positive geometries of generic scalar theories. 
 As discovered in \cite{Jagadale:2022rbl} and briefly discussed in the next section, we further solidify the claim that associahedron acts as a universal polytope, the scattering form on which encodes information about scalar theories containing the most generic interactions.
 
  Now we turn our attention to the 
planar massless $\phi^{3}$ theory
and we will show how scattering amplitude of the theory can be realized in terms of combinatorial polytope in the kinematic space. 
In view of this, first we rewrite the kinematic space of scattering momenta of $n$-particle scattering amplitudes in which all the external states are scalars and denoted  by ${\cal K}_{n}$ \cite{Arkani-Hamed:2017mur}. 
For the case $D > n$ dimensions, the kinematic space ${\cal K}_{n}$ is given by linearly independent planar variables $X_{ij}$ with $|i-j|\geq 2$. That is, the set $\{\, X_{ij}\, \vert\, \vert i - j\vert\, \geq\, 2\, \}$ spans ${\cal K}_{n}$. There is a one-to-one correspondence between these planar variables $X_{ij}$ and the diagonals of an $n$-gon
that maps each triangulation to a Feynman diagram of planar $\phi^{3}$ theory and provides us a convenient way to introduce the combinatorics of pole structure of the S-matrix. 

In order to yield the geometric realization, first consider a positive wedge whose boundaries correspond to the poles of the scattering amplitudes. 
This wedge is given by the expression 
\begin{equation}
\label{positive_quadrant}
    X_{ij} \geq 0 \ ;  \hspace{1cm} \forall (i,j) \text{ such that } |i-j|>1 \ .
\end{equation}
Now we consider a set of hyperplanes that will capture the combinatorics of the poles.  
For any triangulation $T$ consider the intersection of set of hyper-planes given by, 
\begin{equation}
\label{hyperplane_eqn}
    X_{i \, j} + X_{i+1 \, j+1}-X_{i \, j+1}-X_{i+1 \, j} = c_{i \, j} \ ;  \hspace{ 1cm } \text{ for all } (i,j) \notin T^{c} \ ,
\end{equation}
where $T^{c}$ is the triangulation that is obtained by rotating $T$ by $\frac{2 \pi }{n}$ in counter-clockwise direction. 
The intersection of the positive wedge and these hyper-planes leads ABHY realisation of associahedron. 
A canonical convex realisation of the associahedron in the positive quadrant of the kinematic space $\mathcal K_n$ was given in \cite{Aneesh:2019cvt}.

Associahedron is a {\it simple polytope} and there is a projective form associated with it given by 

\begin{equation}
\label{canonical_form}
\Omega(\mathcal M_n) = \sum_{{\text{vertex}} \ Z} {\text{sign}} (Z) \wedge_{a=1}^{n-3} d \log X_{i_a,j_a} \ , 
\end{equation}
Here the sum runs over vertices of the associahedron which correspond to complete triangulations of $n$-gon, and the wedge product is over diagonals of those complete triangulations. The sign$(Z)$ is evaluted on the ordering of the facets in the Wedge product.

Moreover, for each ordering of the facets, sign$(Z) \in \{\pm 1\}$ denotes its orientation relative to the inherited orientation. This form when restricted to the ABHY realisation of associahedron in the kinematic space provides with the scattering amplitude of the planar $\phi^{3}$ theory. 

Considering the sign flip rule between 
\begin{equation}
 {\text{sign}} (Z) = - {\text{sign}} (Z^\prime) \ . 
\end{equation}
Now one needs to restrict the form in \eqref{canonical_form} to the subspace defined by the set of hyperplanes \eqref{hyperplane_eqn}. It follows that an overall volume factor can be extracted from the wedge product of d$log$ forms and we define it as,
\begin{equation}
d^{n-3}X:= {\text{sign}} (Z) \wedge_{a=1}^{n-3} dX_{i_a,j_a}
\end{equation}

Then $\Omega(\mathcal M_n)$ takes the form
\begin{equation}
\Omega(\mathcal M_n) =\Big( \sum_{{\text{planar}} \ Z} \frac{1}{\prod_{a=1}^{n-3}X_{i_a,j_a}} \Big) d^{n-3}X = \mathcal A_n d^{n-3}X \ . 
\end{equation}
where $\mathcal A_n$ gives the expected amplitude.

It is worthwhile to mention that, in literature the expressions inside the parentheses is defined as ``canonical rational function" ($\underline{\Omega}(\mathcal M_n))$\cite{Arkani-Hamed:2017tmz}.
Therefore, one can say the expected amplitude is same as the canonical rational function with the expression 
\begin{equation}
\underline{\Omega}(\mathcal M_n) = \mathcal A_n \ .
\end{equation}

For example $n=6$, $\Omega_{n=6}^{(3)}$ we yield 
\begin{eqnarray}
\label{form6}
\Omega_{n=6}^{(3)} &=& d \log \frac{X_{2,4}}{X_{1,3}} \wedge  d \log \frac{X_{1,4}}{X_{4,6}} \wedge \frac{X_{1,5}}{X_{4,6}} \cr
&& \cr
&+& d \log \frac{X_{2,6}}{X_{1,3}} \wedge  d \log \frac{X_{3,6}}{X_{1,3}} \wedge \frac{X_{4,6}}{X_{3,5}} \cr
&& \cr
&-& d \log \frac{X_{2,6}}{X_{1,5}} \wedge  d \log \frac{X_{2,5}}{X_{3,5}} \wedge \frac{X_{2,4}}{X_{3,5}} \cr
&& \cr
&-& d \log \frac{X_{2,4}}{X_{1,3}} \wedge  d \log \frac{X_{4,6}}{X_{3,5}} \wedge \frac{X_{2,6}}{X_{1,5}} \ . 
\end{eqnarray}
For this case the set of equations among the kinematic variables that locate the associahedron in the positive quadrant of the kinematic space are given by, 
\begin{align*}
\{ X_{13},& X_{14} ,X_{15} \} \ge 0\, \cr
&&\cr
X_{26} &=  -X_{13} + c_{26}  \ge 0\ ,  \cr
&& \cr
X_{36} &=  - X_{14} + c_{36} \ge 0\ ,  \cr
&& \cr
X_{46} &= -X_{15} + c_{46} \ge 0 \ , \cr
&& \cr
X_{25} &=  X_{15} -  X_{13} + c_{25} \ge 0\ , \cr
&& \cr
X_{35} &=  X_{15} -  X_{14}+ c_{35} \ge 0 \ , \cr
&& \cr
X_{24} &=   X_{14} -  X_{13} + c_{24} \ge 0 \ . 
\end{align*}
When the canonical form in \eqref{form6} is pulled back to the geometry defined by the above equations, we recover  corresponding six point amplitude (multiplying the volume form $\underbrace{dX_{13} \wedge dX_{14}\wedge dX_{15}}$),
\begin{eqnarray}
\mathcal A_6 &=& \frac{1}{ X_{13}  X_{14}  X_{15}} +  \frac{1}{X_{13}  X_{15} X_{35}} +  \frac{1}{X_{15} X_{25} X_{35}} +  \frac{1}{X_{25} X_{15} X_{24}}\cr
&& \cr
&& +  \frac{1}{X_{15} X_{14} X_{24}} 
+  \frac{1}{ X_{25}  X_{26}  X_{35}} +  \frac{1}{ X_{25}  X_{24} X_{26}} +  \frac{1}{X_{13} X_{35} X_{36}}\cr
&&\cr
&&+  \frac{1}{X_{35} X_{36} X_{26}} +  \frac{1}{X_{13} X_{36} X_{46}} 
+  \frac{1}{X_{36} X_{46} X_{26}} +  \frac{1}{X_{26} X_{46} X_{24}} \cr
&& \cr
&&+  \frac{1}{ X_{14} X_{46} X_{24}} +  \frac{1}{ X_{14} X_{46} X_{13}} \ . 
\end{eqnarray}

One of the important property that the associahedron has is the factorization property. This property has direct mapping to the factorization property of the factorization of the scattering amplitude when a propagator is taken onshell. We will discuss it in some details and will be highly using it for the derivation of the recursion relations.\\

{\bf Factorization property}: In the following we briefly review the factorization properties of the amplitude based on the construction discussed in \cite{Arkani-Hamed:2017mur}. 
It turns out the associahedron factorizes combinatorially that means each facet is combinatorially identical to a product of two lower dimensional associahedra. 
Further it is shown that the polytope factorizes geometrically and the geometric factorization directly implies amplitude factorization.
Therefore one can say the locality and unitarity of the scattering amplitude (these spacetime features of the underlying theory culminate in the form of factorization in the scattering amplitude) can be read as emergent properties of the geometry.
Now we discuss the amplitude factorization in the context of associahedron. 
The residue of the canonical form for the facet $X_{i,j}=0$ factors:
\begin{equation}
{\text{Res}}_{X_{ij=0}} \Omega(\mathcal M_n) = \Omega(\mathcal M_L) \times \Omega(\mathcal M_R) \ . 
\end{equation}
Here $\mathcal M_L$ and $\mathcal M_R$ known as left associahedron and right associahedron respectively and reside in independent kinematic spaces. 
The above expression provides the factorization of the amplitude.
 For an example, we consider $n=6$ associahedron lives in kinematic space. 
Consider the facet $X_{2,5}=0$, which is geometrically a quadrilateral is a product of two 4-point associahedra, {\it i.e.} two line segments.
The same analysis also holds for the facets $X_{1,4}=0$ and $X_{3,6}=0$.
But all the other facets of this associahedron fig. are pentagon and are product of a pentagon and a point which are associahedrons for 5-point and three point amplitudes respectively.

\subsection{Deformed realization of the associahedron}

As already mentioned above, the associahedron carry the same information of the scattering data as contained in the planar Feynman diagrams for a given $n$-point amplitude. The associahedron is located in the kinematic space by the set of constraint equations \ref{hyperplane_eqn} which defines a set of hyperplanes within the positive quadrant of $\mathcal K_n$.

ABHY realisation can be read as realisation of the combinational associahedron in an abstract embedding $\mathbb R^{\frac{n(n-3)}{2}}$.
Considering this realisation which provides $\phi^3$ amplitude is given by identification of this embedding space along with the kinematic space whose unit vectors in $\mathbb R^{\frac{n(n-3)}{2}}$ identified with $X_{ij}$.
For the massless scalar theory the poles of scattering amplitude are located at $X_{ij}=0$ and for massive case it generalised to 
\begin{equation}
  \widetilde X_{ij}= X_{ij}-m^2 \geq 0 \ ,
\end{equation}
for $|i-j|>1$.
The boundary of the associahedron now moves to $X_{ij}=m^2$.

The hyperplane equations for the massive case can be schematically represented as $\widetilde X_{ij} +\widetilde X_{k\ell } =\widetilde X_{i \ell}+\widetilde X_{kj}+ c_{ijk \ell}$ with $i< k< j<l$, equivalently it says that both $\widetilde X_{ij}$ and $\widetilde X_{kl}$ propagators cannot appear in the same Feynman diagram given that all $\widetilde X_{ij} \ge 0 $ and $c_{ijkl} > 0$. However, the same combinatorics would be captured if we had instead consider hyper-planes  given by the following equations,
\begin{equation}
\label{deformed_hyperplane}
    \alpha_{ij} \widetilde X_{ij} + \alpha_{k \ell} \widetilde X_{k \ell } - \alpha_{i\ell} \widetilde X_{i\ell} - \alpha_{k j} \widetilde X_{k j } = c_{ijk\ell} \ .
\end{equation}
With $\alpha_{ij},\alpha_{k\ell}, \alpha_{i \ell} $ and $\alpha_{k j}$  being positive. Heuristically, one could argue that since the pole $\frac{1}{X_{ij}}$ is ``no different" than $\frac{1}{X_{k \ell}}$ or any other pole, all the $\alpha$ factors should be same and the over all positive $\alpha$ can be absorbed into $c_{ijk\ell}$. However, the question arise: What happens when we take the $\alpha$s to be different? And the answer is that taking $\alpha$s to be different provides us with a realisation of associahedron that gives the scattering amplitude of a theory with multiple scalar fields \cite{Jagadale:2022rbl}. 

To deal with the massive propagators together with the $\alpha$-deformation of the kinematic variables in one shot, we define a new kinematic variable,

\begin{equation}
\label{shifted_wedge}
    \kappa_{ij} := \alpha_{ij} \widetilde X_{ij} \ . 
\end{equation}
This redefinition modifies the set of hyperplane equations accordingly. Now our goal is to locate the ABHY associahedron inside the space whose coordinates are the $\kappa_{ij}$ variables and that will eventually lead to exact same structure of the undeformed associahedron. To make a parallel to the massless case, if one identifies the embedding space unit vectors with $\kappa_{ij}$ then we get the above realisation. But these deformed space are not isometric to the kinematic space $\mathcal K_n$\footnote{Only if one chooses all the $\alpha_{ij}$s and $m_{ij}$s to be equal, these two spaces become isometric.}. From the perspective of original kinematic space the set of hyperplanes \eqref{deformed_hyperplane} cut out a deformed realisation of the associahedron. Here deformed strands for it is stretched or compressed in some directions denoted by $\mathcal M_{n}^{\{\alpha\}}$.
For $\kappa_{ij}\geq 0$, the positive edge of the geometry gets shifted. 
On a similar note, we get geometry with rotated hyper-planes for the constraint relation \eqref{deformed_hyperplane}. In this paper, we will mostly resort to the deformed realisation of the associahedron where the associahedron boundaries remain at same $\widetilde X_{ij}=0$ as the undeformed case but the associahedron is located in the kinematic space $\mathcal K_n^+$ by using the deformed hyperplane equations \eqref{deformed_hyperplane}\footnote{In the geometric picture, the modifications \eqref{shifted_wedge} and \eqref{deformed_hyperplane} correspond to the shift of the positive wedge and the rotation of the hyperplanes compared to the undeformed case.}. And this fact alone leads to the following scattering form on the associahedron,

\begin{flalign}\label{cfodass2}
 \Omega_{n}(\widetilde X_{ij})\, =:\, \mathcal A_{n}\, \wedge_{(ij)\, \in\, T_{0}}\, d^{n-3}\tilde{X}_{ij} \ , 
\end{flalign}
where 
\begin{flalign}\label{cfodass3}
\mathcal A_{n}\, =\, \sum_{T}\, (\, \frac{\prod_{(mn)\, \in\, T_{0}}\alpha_{mn}}{\prod_{(kl)\, \in\, T}\alpha_{kl}}\, \prod_{(mn)\, \in\, T}\, \frac{1}{\tilde{X}_{mn}}\, ) \ .
\end{flalign}
We denote the initial triangualtion as $T_0$ (which serves as the basis variable for realizing the associahedron) and the other triangualtions as $T$.

$\mathcal A_{n}$ is called the deformed amplitude, which is related to the $n$-point scattering amplitude of a theory with multiple scalar fields with generic cubic interactions. The numerator factor is common to all the terms in \eqref{cfodass3} and can be stripped off to define something called {\it stripped amplitude} $\mathcal A^{\prime}_{n}$ with the expression

\begin{flalign}\label{cfodass1}
\mathcal A^{\prime}_{n}\, =\, \sum_{T}\, (\, \frac{1}{\prod_{(kl)\, \in\, T}\alpha_{kl}}\, \prod_{(mn)\, \in\, T}\, \frac{1}{\tilde{X}_{mn}}\, ) \ . 
\end{flalign}

For completeness, we note that the canonical form \eqref{cfodass2} on the deformed associahedra generate $n$-point amplitude $\mathcal A_n$ in a sequence of QFTs having the Lagrangian

\begin{flalign}
\label{int_lag}
L_{n}(\phi_{1},\, \phi_{2},\, \dots,\, \phi_{\rfloor\, \frac{n}{2}\, \rfloor}\, )\, =  \frac{1}{2} \sum_{I=1}^{\frac{n}{2}} {\text{Tr}} \big(\partial_\mu \phi_i \  . \ \partial_\mu \phi_i \big) - \sum_{1 \leq I \leq J \leq K \leq \frac{n}{2} \ , \ I+J=K \  {\text{mod}} \ n}
\frac{\lambda_{IJK}}{3!} \phi_I \phi_J \phi_K \ . 
\end{flalign}
where $\phi_{1}$ is a massless bi-adjoint scalar and $\phi_{I}, I\, >\, 1$ are massive scalars with distinct masses $m_{I}\, \neq\, 0$  barring the fact the couplings $\lambda_{IJK}$ s satisfy a set of appropriate non-linear relations with deformation parameters $\alpha_{ij}$ s.

For an example, in the following we will discuss the mixed scalar field amplitude with fine-tuned couplings. 
We consider the generalisation of the analysis where all the external particles are massless.
However, all the propagators have a pole at $m^2$.
In this picture the deformed kinematic variables are given by
\begin{eqnarray}
\alpha_{ij} &=& 1 \  \forall \ (ij) \ , \cr
\kappa_{ij} &=& \tilde{X}_{ij} = X_{ij}-m^2 \  {\text{if}}  \ |j-i| \geq 2 \ , \cr
\kappa_{i,i+1} &=& \kappa_{1n}= 0 \ . 
\end{eqnarray}
%
%
As stated, here all external particles are massless, therefore the kinematic space $\mathcal K_n$ is co-ordinated by $X_{ij}$ variables along with the convex realisation
\begin{eqnarray}
 \kappa_{ij}+\kappa_{i+1,j+1}-\kappa_{i,j+1}-\kappa_{i+1,j}= c_{ij} \ . 
\end{eqnarray}
The above relation implies the ABHY associahedron in the positive quadrant of the kinematic space $\mathcal K_n^+$ with facets at $\kappa_{ij}=0$ or $X_{ij}=m^2$.
The canonical form on the associahedron realised through 
\begin{eqnarray}
 \Omega_n = \sum_T \prod_{(kl)\in T} \frac{1}{\kappa_{kl}} \wedge_{(ij)\in T_0} \ d^{n-3} \kappa_{ij} \ . 
\end{eqnarray}
The above form provides tree-level $n$-point amplitude for a scalar field theory containing two fields $\phi_1 , \phi_2$. 
Here $\phi_1$ is massless and $\phi_2$ has mass $m$.
The interaction term in Lagrangian is given by $\phi_2^3+\phi_1^2\phi_2+\phi_1\phi_2^2$.
The interactions terms share the same coupling strength.

\subsection{Interaction couplings and deformation parameters} 

In this section we briefly discuss the relation between the coupling constants of the interaction terms present in the Lagrangian with the deformed amplitude (in particular with deformation parameters). 
 
We take the example of single parameter deformation,
\begin{align}
    \alpha_{ij}=1~ \forall~ |i-j|=2 ~\text{mod}~ n \ , \cr
    \alpha_{ij}=\alpha ~ \text{otherwise} \ , 
\end{align}
and consider the six-point deformed associahedron. It is straightforward to see the corresponding canonical form gives  the $n=6$ colored ordered tree-level amplitude in a theory  having the interaction term \cite{Jagadale:2022rbl}
\begin{flalign}\label{tsig}
V(\phi_{1}, \phi_{2})\, =\, \lambda_{1}\, \phi_{1}^{2}\, \phi_{2}\, +\, \lambda_{2}\, \phi_{1}\phi_{2}^{2}\, +\, \lambda_{3}\phi_{2}^{3} \ , 
\end{flalign}
where $\phi_{1}$ is a massless scalar field and $\phi_{2}$ is a massive scalar field with mass $m$. 

In this picture, the mapping of the coupling constants with the deformation parameter takes the form
\begin{flalign}
\label{coupling_rel}
\alpha\, =\, \frac{\lambda_{1}\lambda_{3}}{\lambda_{2}^{2}} \ . 
\end{flalign}
Note that the deformation parameter $\alpha$ has a non-linear dependence on the coupling constant $\lambda$.

Generalisation to an arbitrary number of external particles having an interaction potential as in \eqref{int_lag} is still not completely settled \footnote{Consider an $n$-gon that we triangulate using internal chords. The connection to scattering process is made through each triangulation is dual to a Feynman diagram with every triangle being dual to a cubic vertex.}. The situation becomes more complicated as the number of particles increases, the coupling constants need to satisfy some relation among them before it can be non-linearly related to the deformation parameters, a more detailed discussion on this topic can be found in \cite{Jagadale:2022rbl}.

\section{A specific case of recursion} 
\label{recursion}

We try to generalize the recursion relation of \cite{He:2018svj,Yang:2019esm} to the more general case of deformed associahedron which are shown to be the positive geometry of massive multi scalar field amplitudes having cubic interactions of different strengths among them. We start by reviewing the construction of these recursion relations and consequently adapting it to the case of interest here.

A deformed $d$-dimensional polytope $A$ in kinematic space is described by a basis of planar kinematic variables $X_{A_{i}}$ where $ i=1,\ldots,d$. The other kinematic variables are expressed in terms of these basis variables. We adopt the notation $\widetilde X_{ij}=: (X_{ij} -m_{ij}^2)$ to deal with the massive case. In the massless channels $m_{ij} =0$.

Let us now consider the following one-parameter rescaling of a subset of basis variables :
\begin{align}
\label{shift}
\widetilde X_{A_i} \to z\,\widetilde X_{A_i} \ ; \quad\quad i=1,2,\ldots,k\quad\text{and}\quad k\le d \ . 
\end{align}

Under this rescaling of basis variables, a subset of the dependent variables are also modified. We denote the modified $n$-point amplitude as $\mathcal A_n(z\widetilde X,C)$

%
Let us now consider the following integral :
\begin{align}
\label{integrand}
\oint dz \frac{z^k}{z-1} \mathcal A_n(z\widetilde X,C) \ . 
\end{align}
By construction, the original amplitude is given by the residue of the integrand at $z=1$. By Cauchy's residue theorem, the contour integral can be expressed as the sum of residues from finite poles and (possibly) a pole at infinity :
\begin{align}
\label{Cauchy}
\mathcal A_n(X,C)=-\left(\text{Res}_{\infty} +\sum_{\text{finite\,poles}} \text{Res}_{z_i}\right)\,\left(\frac{z^k}{z-1} \mathcal A_n(z\widetilde X,C)\right) \ . 
\end{align}
We soon show that the integrand in \eqref{integrand} has no pole at $z=0$ or $z=\infty$.

{\it The large z behavior of the amplitude of a theory is unaffected by inclusion of more than one type of scalar particles, so that takes care of one part for the recursion relation to go through, whereas the other property needed for the recursion relation to hold is the factorization property of the scattering amplitude. And  as we will see, that is modified in a specific manner for the recursion relation to work for the case with more than one type of particles.}

Whereas a canonical convex realisation \eqref{hyperplane_eqn} of the associahedron in kinematic space is already given in \cite{Arkani-Hamed:2017mur}. The general form of these equations are,

\begin{align}
\label{compatibleX}
\widetilde X_{B_i}= \frac{1}{\alpha_{B_i}}\left(\epsilon_{B_i}+\alpha_i\widetilde X_i+ \sum_{j=1}^k \lambda_{ij}\alpha_{A_j} \widetilde X_{A_j}\right) \ , 
\end{align}
where $\widetilde X_{B_i}$ denote planar variables compatible with the basis variables, $\epsilon_{B_i}$ and $X_i$ denote respectively suitable linear combinations of constants that appear in the convex realisation and of the unmodified basis variables $\{X_{A_j}\}_{ j=k+1,...,d}$ ,  $\lambda_{ij}$ are real constants.

The poles of the modified amplitude $\mathcal A_n(z X,C)$ are given by solutions of equations $\widehat {\widetilde X}_{B_i}$ where $\widehat {\widetilde X}_{B_i}$ denotes the modified $\widetilde X_{B_i}$ under the shift \eqref{shift}.\\
%
At finite physical poles, the amplitude factorizes as :
\begin{equation}
\label{factorization}
\text{limit}_{\widehat {\widetilde X}_{B_i} \to 0}~ \alpha_{B_i}\widehat {\widetilde X}_{B_i}(z) \mathcal A_n(z \widetilde X,C) = \mathcal A_{B_i}^{\text{limit}}(z_{B_i}\widetilde X,C) \ , 
\end{equation}
$A_{B_i}^{\text{limit}}$ is given by the product of lower point amplitudes.
For example, in the case of tree amplitudes \eqref{factorization} takes the form:

\begin{equation}
\label{factor_deform}
\text{limit}_{\widetilde X_{pq} \to 0} \, \widetilde X_{pq} \,\mathcal A_{1,2,...,n}(\widetilde X,C)= \frac{\alpha_{mk}}{\alpha_{pq}} \mathcal A_{p,...,q-1,I} \times \mathcal A_{I,q,...,p-1} \ , 
\end{equation}

with $I$, the internal particle going on-shell and $(mk) \in T_0$. $\mathcal A$ is the deformed lower point amplitude of the form \eqref{cfodass3}. Note that one of the $\alpha$ factor in the numerator cannont be absorbed in the lower point amplitudes for reason explained below. Rest of the numerator factors nicely get distributed among the two lower point functions. The reason being, from the geometry side, taking one of the propagator on-shell amounts to approaching a specific facet of the the associahedron. Now as we know each facet of the assocaihedron can be factorized as the product of lower dimensional associahedrons and the canonical form on that facet are products of canonical forms of the lower dimensional associahedrons which provides us with the required factorization property for the recursion to work. Although one should note one crucial thing, that is the numerator $\alpha$ factors in the deformed amplitudes come soley from the volume form or the wedge product. Now the volume form for the co-dimension one facet is always one dimension less. That amounts to the canonical form of the codimension one facet which is given by product of lower dimensional associahedrons,  will  be able to accommodate all but one (appropriately chosen) of the $\alpha$ factors of the numerator, inside the product of lower point deformed amplitudes. It is worth mentioning, the deformed geometry although deformed inside the original kinematic space is still an associahedron, and any facet of the associahedron being product of lower dimensioanl associahedrons is suppopsed to be capturing deformed amplitudes as is rightly written in \eqref{factorization}, a fact we will demonstrate in Appendix \ref{volume_cal} with few simple examples. This is the key difference compared to the recursion relation for undeformed associahedron, the canonical form of the deformed associahedron factorizes only in this specific manner when one of the numerator factor is taken out mutipying the factorized amplitiude.  \\

Since keeping track of the numerators factors is a bit difficult if we use the above factorization formula, we can rewrite it as
\begin{equation}
\text{limit}_{\widetilde X_{pq} \to 0} \, \widetilde X_{pq} \,\mathcal A_{1,2,...,n}(\widetilde X,C)= \frac{\prod_{(mk)\in T_0}\alpha_{mk}}{\alpha_{pq}} \mathcal A'_{p,...,q-1,I} \times \mathcal A'_{I,q,...,p-1} \ , 
\end{equation}

$\mathcal A'$ is the deformed lower point amplitude without the numerator factors {\it i.e.} the stripped amplitude of the form \eqref{cfodass3}\footnote{It is still a recursion relation as we do build the higher point amplitude from some lower point ones.}. The reason to work with the stripped amplitude is that once we strip off the numerator factor the factorization property is manifested very easily as we can view the stripped amplitude is captured by the associahedron when it is relaized in $\kappa_{ij}$ space where the associahedron remains undeformed in shape. We will be explicitly using this stripped amplitudes for factorization purpose from now on keeping in mind what the original physical picture is as in \eqref{factor_deform}. From the geometric point of view, these numerator factors are directly related to the volume of the polytope. They act as a scale factor to the original undeformed volume of the polytope when the deformed polytope is located in the same coordinate system. Note that, we distribute the $\alpha$ factors in the denominator among the lower point functions when constructing the recursion relation. Depending upon which propagator goes on-shell, the higher point amplitudes can be factorized to different lower point amplitudes. This valuable information is encoded in the recursion relation in such a manner the $\alpha_{ij}$ s in the  denominator are distributed among the lower point amplitudes using factorization. This is in contrast to the usual $\phi^3$ theory with a single field where only one type of lower point amplitude is present. \\
The residue of the integrand then reads,
\begin{equation}
\text{Res}_{\widehat{\widetilde X}_{B_i}} \frac{z^k}{z-1}A_n(z \widetilde X,C)=\frac{z_i^k}{(\sum_{j=1}^k\lambda_{ij} \widetilde X_{A_j})(z_i-1)}\mathcal A_{B_i}^{\text{limit}}(z_{B_i} \widetilde X,C) \ . 
\end{equation}
This leads to the recursion formula for $n$-point deformed amplitude with a given initial triangulation,
\begin{align}
\label{generalscalingformula}
A_n( \widetilde X,C)=\sum_{B_i}\frac{z_i^k}{ \widetilde X_{B_i}}A_{B_i}^{\text{limit}}(z_{B_i} \widetilde X,C) \ .
 \end{align}

 \subsection*{Proof: No pole at $z=0$ }
 The canonical function of simple polytopes is given by \cite{Salvatori:2019phs},
\begin{equation}
\label{canonical_function}
\underline{\Omega_p} = \sum_{v\,\in\,\text{vertices}}\,\,\prod_{\text{facets}\,f\,\in\,v} \frac{1}{(\bm{\alpha. \widetilde X})_f} \ , 
\end{equation}
where we introduced the compact notation $ (\bm{\alpha.\widetilde X})_i =\prod_i \alpha_i \widetilde X_i$.

For a $d$ dimensional simple polytope,  exactly $d$ facets are adjacent to any vertex $v$. Thus, the product $\prod \frac{1}{\widetilde X_f}$ is proportional to $\frac{1}{\widetilde X^d}$. Once $k$ of the basis variables are rescaled, among the $d$ $\widetilde X_{ij}$'s that appear in the product in \eqref{canonical_function} at most $k$ of them have a $z$ dependence that can lead to a pole at $z=0$. We do not get a pole at $z=0$ arising from non-basis variables. Thus the product in \eqref{canonical_function} can at most lead to an order $k$ pole at $z=0$ and that contribution is precisely cancelled by the $z^k$ factor in the integrand \eqref{integrand}. Hence, it is proved that the integrand in \eqref{integrand} has no pole at $z=0$.

  \subsection*{Proof: No pole at $z \rightarrow \infty$ }

  To obtain the $O(\frac{1}{z^k})$ contribution, we need $k$ of the $\widetilde X_{ij}$ in the denominator of \eqref{canonical_function} shifted and $(d-k)$ of them unaffected. If we denote the unaffected ones as $ X_j^{\circ}$, we have the following contribution to the amplitude :
\begin{equation}
\left(\sum_{\text{vertices shared by}\,(d-k)\,X_j^{\circ}} \frac{1}{\prod_{k}( \bm{\alpha \cdot \widetilde X})_k}\right) \frac{1}{\prod_{j=1}^{d-k} (\bm{\alpha \cdot X_j^{\circ}})} \ . 
\end{equation}
The quantity in the brackets is the canonical function for the ABHY polytope when all the $ X_j^{\circ} \to 0$, let us  denote it by $A_k(\widetilde X,C)$. Since all the $\widetilde X$'s in $A_k$ are shifted, as $z \to \infty$  the leading contribution comes from setting $C \to 0$ :
\begin{equation}
\text{limit}_{z \to \infty} A_k(z\widetilde X,C) \sim \frac{1}{z^k} A_k(\widetilde X,0)=0 \ . 
\end{equation} 
This fact can be argued in a very simple manner, as $A_k(\widetilde X,0)$ is always given by the product of lower point functions by factorization which can be explicitly shown to vanish e.g. $n=4$ cubic deformed amplitude with $T_0=(13)$ is given by $\mathcal A_4=\alpha_{13}(\frac{1}{\alpha_{13}X_{13}}+\frac{1}{-\alpha_{13}X_{13}+C})$ which goes to zero as $C \to 0$. Geometrically, it implies that the deformed associahedra have shrunk to zero size.

With the proof of the integrand in \eqref{integrand} has no pole at infinity, our derivation of the recursion relation for the general case of deformed associahedra is completed.

{\it One of the main motives of all these series of papers is that the same recursion can be used to obtain the scattering amplitude of a family of theories with scalar particles having a most general kind of interaction among them. This paper takes a major leap in this direction to extend the recursion from a single type of particle to a theory with more than one type of particles interacting via. cubic couplings of different strengths. Note that although this recursion relation is mathematically simliar in structure to the ones given in \cite{He:2018svj,Yang:2019esm,Kojima:2020tox,John:2020jww}, there are profound physical implications that emerge out of these relations in the case of deformed associahedron which we will see in later sections.}

In the following we will explicitly workout some examples to show that the recursion relations lead the correct amplitude.
In the following, first we consider four-point and six point tree amplitude in massless $\phi^3$ theory, afterwards we consider loop amplitude 
as well.

\subsection{ Example I : Four-point amplitude in massless \texorpdfstring{$\phi^3$} ~ theory}

To begin with, first we consider tree level four-point amplitude in the context of massless $\phi^3$ theory.
For this case the basis planar variables are given by $s$ and $t$ namely {\it{Mandelstam variables}}.
The corresponding deformed realization of the associahedron is given by the equation $\alpha s +\beta t= c$, 
where $c$ is a constant. 
Out of the two basis variables, we consider the variable $s$ and rescale it as $s \rightarrow zs$.
It is straightforward to see that the recursion relation gives only one term 
\begin{equation} \label{m4}
    \mathcal A_4= \frac{z_t}{t}\times \frac{\alpha}{\beta} \ , 
\end{equation}
where $z_t = \frac{c}{\alpha s}$.
Moreover, plugging $z_t$ in the \eqref{m4} we yield the four-point deformed amplitude in massless $\phi^3$ theory 
\begin{equation} \label{m42}
    \mathcal A_4= \frac{c  \alpha}{\alpha \beta s t} =  \frac{\alpha s + \beta t}{\alpha \beta} \frac{\alpha}{st}  = \alpha\Big(\frac{1}{\alpha s} + \frac{1}{\beta t}\Big) \ . 
\end{equation}

{\it The four point deformed amplitude has only one $\alpha$ factor in the numerator. Now a four point amplitude has only one propagator and while taking it on-shell, the amplitude factorizes into two three-point functions. The single $\alpha$ factor could have been contributed to one of the three point sub amplitude whereas the other one remains to be unity. This is compatible due to the presence of more than one type of three point vertices in the Lagrangian with different coupling constants. But this scheme of assigning all the numerator factors to the lower point amplitudes runs into inconsistencies as we move to the higher point amplitudes.}

\subsection{Example II : Six point deformed amplitude in massless \texorpdfstring{$\phi^3$}~ theory}\label{six_point_recursion}

As an another example now we consider 
 six point amplitude in deformed case. 
 To begin with first we write down the planar variables under the deformation
\begin{eqnarray}
X_{26} &=& \frac{1}{\alpha_{26}} \big(-\alpha_{13}X_{13} + c_{26}\big) \ ,  \cr
&& \cr
X_{36} &=&  \frac{1}{\alpha_{36}} \big(-\alpha_{14} X_{14} + c_{36}\big) \ ,  \cr
&& \cr
X_{46} &=& \frac{1}{\alpha_{46}} \big(-\alpha_{15}X_{15} + c_{46}\big) \ , \cr
&& \cr
X_{25} &=& \frac{1}{\alpha_{25}} \big(\alpha_{15} X_{15} - \alpha_{13} X_{13} + c_{25}\big) \ , \cr
&& \cr
X_{35} &=& \frac{1}{\alpha_{35}} \big(\alpha_{15} X_{15} - \alpha_{14} X_{14}+ c_{35}\big) \ , \cr
&& \cr
X_{24} &=&  \frac{1}{\alpha_{24}} \big(\alpha_{14} X_{14} - \alpha_{13} X_{13} + c_{24}\big) \ . 
\end{eqnarray}
The geomtric relazation of this assciahedron is given in Fig.\ref{fig:6point_asso}. In this case we have six basis variables. 
Out of them we consider the variables $X_{13}$ and $X_{14}$ and rescale them 
\begin{eqnarray}
X_{13} \rightarrow z X_{13} \ ;  \  X_{14} \rightarrow z X_{14} \ . 
\end{eqnarray}
It is very straightforward to see that under the above rescaling, apart from $X_{46}$ all dependent variables will be deformed and we have following 
recursion relations. 
In this case we have five recursion terms which we denote by $\mathcal R_i^d \ ; \ i =1,2,..5$. Out of five recursion terms, for illustration in the following we will present 
only two of them, 
\begin{eqnarray}
\label{second6point}
\mathcal R_2^d = \frac{c_{26} c_{36} \big(c_{35} - c_{36} + c_{46}\big)}{\alpha_{13}X_{13} \alpha_{36} X_{36}  \alpha_{46}X_{46} \big(c_{36}  \alpha_{13}X_{13} - c_{26}  \alpha_{14} X_{14}\big) \big(c_{36}-  \alpha_{15}X_{15} - c_{35}\big)} \ , 
\end{eqnarray}

{\tiny
\begin{eqnarray}
\label{6point_quadraticpole}
\mathcal R_3^d &=& \frac{\big(c_{26}-c_{25}\big) \big(c_{24}-c_{25} +c_{35}\big) \big(c_{25} + \alpha_{15} X_{15}\big)^2}{\alpha_{15} X_{15} \alpha_{25} X_{25} (\alpha_{25} X_{25}-\alpha_{26} X_{26})  (-\alpha_{14} X_{14} (c_{25} + \alpha_{25}X_{15}) + \alpha_{13}X_{13}(-c_{24} +c_{25}  \alpha_{15}X_{15}))  (-\alpha_{14} X_{14}(c_{25} + \alpha_{15}X_{15} )
+ \alpha_{13}X_{13} (c_{35} +  \alpha_{15}X_{15} )) }.\cr
&&
\end{eqnarray}
}

It is worthwhile to mention that each of the recursion term has one or more spurious poles and they neatly cancel after adding all the terms yielding the stripped amplitude
{\scriptsize
\begin{eqnarray}
\mathcal R^d = \sum_{i=1}^{5} \mathcal R_i^d&=& \frac{1}{\alpha_{13} X_{13} \alpha_{14} X_{14} \alpha_{15} X_{15}} +  \frac{1}{\alpha_{13}X_{13} \alpha_{15} X_{15} \alpha_{35}X_{35}} +  \frac{1}{\alpha_{15}X_{15} \alpha_{25}X_{25} \alpha_{35}X_{35}} \cr
&&+  \frac{1}{\alpha_{25}X_{25} \alpha_{15}X_{15} \alpha_{24}X_{24}} +  \frac{1}{\alpha_{15}X_{15} \alpha_{14}X_{14} \alpha_{24}X_{24}} 
+  \frac{1}{\alpha_{25} X_{25} \alpha_{26} X_{26} \alpha_{35} X_{35}} \cr
&&\cr
&&+  \frac{1}{\alpha_{25} X_{25} \alpha_{24} X_{24} \alpha_{26}X_{26}} +  \frac{1}{\alpha_{13}X_{13} \alpha_{35}X_{35} \alpha_{36}X_{36}} +  \frac{1}{\alpha_{35}X_{35} \alpha_{36}X_{36} \alpha_{26}X_{26}} \cr
&& \cr
&&+  \frac{1}{\alpha_{13}X_{13} \alpha_{36}X_{36} \alpha_{46}X_{46}} 
+  \frac{1}{\alpha_{36}X_{36} \alpha_{46}X_{46} \alpha_{26}X_{26}} +  \frac{1}{\alpha_{26}X_{26} \alpha_{46}X_{46} \alpha_{24}X_{24}} \cr
&& \cr
&&+  \frac{1}{\alpha_{14} X_{14} \alpha_{46}X_{46} \alpha_{24}X_{24}} +  \frac{1}{\alpha_{14} X_{14} \alpha_{46}X_{46} \alpha_{13}X_{13}} \ . 
\end{eqnarray}
}
It is straightforward to see the above expression of $\mathcal R^d$ (after we put back the overall scaling factor of $\underbrace{\alpha_{13}\alpha_{14}\alpha_{15}
}$)  is identical to six-point deformed amplitude in planar massless $\phi^3$ theory\footnote{We want to point out that $\alpha_{ij}$ is not a kinematic variable but a parameter that finally relates to the coupling constants of the theory. The pole structure is still captured by kinematic variables $X_{ij}$ only and having a pole in $X_{ij}$ is equivalent to saying having a pole in $\kappa_{ij} := \alpha_{ij}X_{ij}$ variables. Physically, the poles of the scattering amplitude are given only when a propagator $X_{ij}$ goes on-  shell.}. 

{\it Let's continue our discussion on the assignment of numerator factors to the lower point amplitudes while constructing the recursion relation. As we move to five point amplitude, there are two propagators present. One can fix either of the two to be on-shell, that results into a four point and a three point sub amplitudes. Since the five point deformed amplitudes contains two $\alpha$ factors in the numerator and the four point sub-amplitude contribute only one of the $\alpha$, the remaining one must come from the three point sub-amplitude. But one can show that the same distribution of numerator $\alpha$ factors cannot occur for all the five Feynman diagrams that form the vertices of the associahedron. This is due to one of the three point function is already fixed to be unity from four point factorization. The issue becomes more complicated as we go to the six point amplitude. It can factorize into either two four point functions or a combination of a five point and a three point function depending upon which propagator is taken on-shell. Whereas in the later case leads to a product of three $\alpha$s  in the numerator of six point function, the former case leads to a product of two  $\alpha$s in general. Hence there is a clear inconsistency in assigning the numerator factor to the sub amplitudes. As we go to even higher point amplitudes, the scattering channels into which it can factorize increase and with that the ambiguity of assigning the numerator factor. On the other hand, if we take out one numerator $\alpha$ factor and do not assign any factor to the three point amplitude, one can check for themselves that assigning the remaining numerator factors through the above exercise can be done in a consistent manner.}

\subsection{Example III: Loop in deformed case }

Now we turn our interest to the loop amplitude, in particular the three point deformed one-loop amplitude \footnote{In fact, here we will only deal with the integrand of the loop amplitude, though for notational simplicity it will be denoted as the loop amplitude from now on.}. 
For the loop case, together with $X_{ij}$ we have more kinematic variables denoted with $Y_i$ (to express the relevant kinematic variables, we follow the notation of \cite{Jagadale:2020qfa}).
\begin{eqnarray}
 \widetilde Y_1 &=& \frac{1}{\alpha_{\tilde Y_1} } \big(c_{23} + c_{2L} +c_{3L} - \alpha_{Y_2} Y_2\big) \ ,  \cr
&& \cr
  \widetilde Y_2 &=& \frac{1}{\alpha_{\tilde Y_2} } \big(c_{12} + c_{1L} +c_{2R} - \alpha_{Y_1} Y_1 + \alpha_{Y_2} Y_2 - \alpha_{Y_3} Y_3\big) \ , \cr
&&\cr
 \widetilde Y_3 &=& \frac{1}{\alpha_{\tilde Y_3} } \big(c_{12} + c_{1L} +c_{2L} - \alpha_{Y_1} Y_1 \big) \ , \cr
&&\cr
 X_{2\bar 3} &=& \frac{1}{\alpha_{ X_{2\bar 3}} } \big(c_{12} + c_{1L} -  \alpha_{Y_1} Y_1 + \alpha_{Y_2} Y_2\big)  \ , \cr
&& \cr
 X_{1\bar 2} &=& \frac{1}{\alpha_{ X_{1\bar 2}} } \big(c_{12} + c_{2R} +  \alpha_{Y_2} Y_2 - \alpha_{Y_3} Y_3\big)  \ , \cr
&& \cr
 X_{13} &=& \frac{1}{\alpha_{X_{13}} } \big(c_{23} + c_{2L} -  \alpha_{Y_2} Y_2 + \alpha_{Y_3} Y_3\big)  \ .  
\end{eqnarray}

The one-loop four point stripped amplitude \footnote{The total amplitude is given by multiplying the stripped amplitude with an overall factor of $\underbrace{\alpha_{Y_1}\alpha_{Y_2}\alpha_{Y_3}}$ as we choose $(Y_1,Y_2,Y_3)$ to be the basis variables to express the other kinematic variables.} is given by,
{\scriptsize
\begin{eqnarray}
\label{amp_4loop}
\mathcal A_{{\text{deformed loop}}}  &=& \frac{1}{\alpha_{Y_1} Y_1 \alpha_{Y_2} Y_2 \alpha_{Y_3} Y_3} +  \frac{1}{\alpha_{\tilde Y_1} \tilde Y_1 \alpha_{\tilde Y_2} \tilde Y_2 \alpha_{\tilde Y_3} \tilde Y_3 } +  \frac{1}{ \alpha_{X_{13}} X_{13} \alpha_{ Y_1}  Y_1 \alpha_{ Y_3}  Y_3} 
+  \frac{1}{ \alpha_{X_{13}} X_{13} \alpha_{\tilde Y_1} \tilde Y_1 \alpha_{\tilde Y_3} \tilde Y_3} \cr
&& \cr
&&  + \frac{1}{ \alpha_{X_{13}} X_{13} \alpha_{ Y_1} Y_1 \alpha_{\tilde Y_1} \tilde Y_1} + \frac{1}{ \alpha_{X_{13}} X_{13} \alpha_{ Y_3}  Y_3 \alpha_{\tilde Y_3} \tilde Y_3} + \frac{1}{ \alpha_{X_{1\bar 2}} X_{1\bar 2} \alpha_{ Y_2} Y_2 \alpha_{Y_1} Y_1}+ \ \frac{1}{ \alpha_{X_{1\bar 2}}  X_{1\bar 2} \alpha_{\tilde Y_2} \tilde Y_2 \alpha_{\tilde Y_1} \tilde Y_1} \cr
&& \cr
&&  + \frac{1}{ \alpha_{X_{1\bar 2}}  X_{1\bar 2} \alpha_{ Y_2} Y_2 \alpha_{\tilde Y_2} \tilde Y_2} + \frac{1}{\alpha_{X_{1\bar 2}}  X_{1\bar 2} \alpha_{ Y_1}  Y_1 \alpha_{\tilde Y_1} \tilde Y_1} + \frac{1}{  \alpha_{X_{2\bar 3}}  X_{2\bar 3} \alpha_{ Y_3} Y_3 \alpha_{Y_2}  Y_2}+\ \frac{1}{  \alpha_{X_{2\bar 3}}  X_{2\bar 3} \alpha_{\tilde Y_3} \tilde Y_3 \alpha_{\tilde Y_2} \tilde Y_2}  \cr
&& \cr
&& + \frac{1}{  \alpha_{X_{2\bar 3}}  X_{2\bar 3} \alpha_{ Y_3}  Y_3 \alpha_{\tilde Y_3} \tilde Y_3}+ \ \frac{1}{ \tilde \alpha_{X_{2\bar 3}} X_{2\bar 3} \alpha_{ Y_2}  Y_2 \alpha_{\tilde Y_2} \tilde Y_2} \ . \cr
&& 
\end{eqnarray}
}

Corresponding D-type polytope structure is illustrated in Fig.\ref{geometricy1y2y3}. As like earlier case, we consider two basis variables $Y_1$ and $Y_2$ and rescale them uniformly. 
Then we have the following recursive relations.

For first recursion term we have 
\begin{eqnarray}
\mathcal R_1^{{\text{double deformed loop}}} = \frac{\hat Z_{1}^2 }{\alpha_{\tilde Y_1}\tilde Y_1}  \Bigg[\frac{1}{\tilde {\mathcal Y_{2}} \tilde{\mathcal Y_{3}}} + \frac{1}{\mathcal X_{13} \tilde{\mathcal Y_3}} + \frac{1}{\mathcal X_{13} \hat{Z}_1 \alpha_{Y_1} Y_1} 
+  \frac{1}{\mathcal X_{12} \tilde{\mathcal Y}_2 } + \frac{1}{\mathcal X_{1\bar2} \hat{Z}_1 \alpha_{Y_1} Y_1 } \Bigg] \  , 
\end{eqnarray}
where
\begin{eqnarray}
\hat Z_{1} &=& \frac{c_{23} + c_{2L} + c_{3L}}{\alpha_{Y_2} Y_2} \ ,  \cr
\tilde {\mathcal Y}_{2} &=&  c_{12} + c_{1L} + c_{2R} - \hat Z_{1}  \alpha_{Y_1} Y_1+\hat Z_{1}  \alpha_{Y_2} Y_2 -   \alpha_{Y_3} Y_3                                                                    \ ,  \cr
\tilde {\mathcal Y}_{3} &=& c_{12} + c_{1L} + c_{2L}  -  \hat Z_{1} \alpha_{Y_1} Y_1     \ ,  \cr
\mathcal X_{13} &=& c_{23} + c_{2L} -  \hat Z_{1} \alpha_{Y_2} Y_2 + \alpha_{Y_3} Y_3     \ , \cr
\mathcal X_{1\bar 2} &=& c_{12} + c_{2R} +  \hat Z_{1} \alpha_{Y_2} Y_2 - \alpha_{Y_3} Y_3     \ . 
\end{eqnarray}

Similarly, the second recursion term takes the form 
\begin{eqnarray}
\mathcal R_2^{{\text{double deformed loop}}} =  \frac{\hat Z_{3}^2 }{\alpha_{\tilde Y_3}\tilde Y_3}  \Bigg[\frac{1}{\tilde {\mathcal Y}_{1} \tilde {\mathcal Y}_{2}} + \frac{1}{\mathcal X_{13} \tilde{\mathcal Y}_1} + \frac{1}{\mathcal X_{13} \alpha_{Y_3} Y_3} 
+  \frac{1}{\mathcal X_{2\bar 3} \tilde{\mathcal Y}_2 } + \frac{1}{\mathcal X_{2\bar 3}  \alpha_{Y_3} Y_3 } \Bigg] \ , 
\end{eqnarray}
where
\begin{eqnarray}
\hat Z_{3} &=& \frac{c_{12} + c_{1L} + c_{2L}}{\alpha_{Y_1} Y_1} \ ,  \cr
\tilde {\mathcal Y}_{1} &=& c_{23} + c_{2L} + c_{3L} -\hat Z_{3} \alpha_{Y_2} Y_2   \ ,  \cr
\tilde {\mathcal Y}_{2} &=& c_{12} + c_{1L} + c_{2R} -\hat Z_{3} \alpha_{Y_1} Y_1 + \hat Z_{3} \alpha_{Y_2} Y_2 - \alpha_{Y_3} Y_3  \ ,  \cr
\mathcal X_{13} &=& c_{23} +c_{2L}-  \hat Z_{2} \alpha_{Y_2} Y_2 +\alpha_{Y_3} Y_3 \ .  \cr
\mathcal X_{2\bar3} &=& c_{12} +c_{1L}-  \hat Z_{2} \alpha_{Y_1} Y_1 +\hat Z_{2} \alpha_{Y_2} Y_2 \ . 
\end{eqnarray}

The third recursion term takes the form 
\begin{eqnarray}
\mathcal R_3^{{\text{double deformed loop}}} = \frac{\hat Z_{2}^2 }{\alpha_{\tilde Y_2}\tilde Y_2}  \Bigg[\frac{1}{\tilde {\mathcal Y}_1 \tilde {\mathcal Y}_3} + \frac{1}{ \mathcal X_{1\bar2}  \tilde {\mathcal Y}_1 } + \frac{1}{ \mathcal X_{1\bar2}  \hat Z_{2} \alpha_{Y_2} Y_2 } + 
 \frac{1}{ \mathcal X_{2\bar3}  \tilde {\mathcal Y}_{3}  }+  \frac{1}{ \mathcal X_{2\bar3}  \hat Z_{2} \alpha_{Y_2} Y_2 } \Bigg] \ , 
\end{eqnarray}
where
\begin{eqnarray}
\hat Z_{2} &=& \frac{c_{12} + c_{1L} + c_{2R}-\alpha_{Y_3} Y_3}{\alpha_{Y_1} Y_1 - \alpha_{Y_2} Y_2}   \ ,  \cr
\tilde {\mathcal Y}_{1} &=& c_{23} + c_{2L} + c_{3L} -\hat Z_{2} \alpha_{Y_2} Y_2  \ ,  \cr
\tilde {\mathcal Y}_{3} &=& c_{12} +c_{1L}+c_{2L}-  \hat Z_{2} \alpha_{Y_1} Y_1  \ , \cr
\mathcal X_{1\bar2} &=& c_{12} +c_{1R}+c_{2L}+  \hat Z_{2} \alpha_{Y_2} Y_2 -  \alpha_{Y_3} Y_3 \ , \cr
\mathcal X_{2\bar3} &=& c_{12} +c_{1L} -  \hat Z_{2} \alpha_{Y_1} Y_1 + \hat Z_{2} \alpha_{Y_2} Y_2 \ . 
\end{eqnarray}

The fourth recursion term takes the form 
\begin{eqnarray}
\mathcal R_4^{{\text{double deformed loop}}} = \frac{\hat Z_{23}^2 }{\alpha_{X_{2\bar 3}} X_{2\bar 3}}  \Bigg[\frac{1}{\alpha_{Y_3} Y_3 \hat Z_{23} \alpha_{Y_2} Y_2} + \frac{1}{\tilde {\mathcal Y}_3 \tilde {\mathcal Y}_2} + \frac{1}{\tilde {\mathcal Y}_3 \alpha_{Y_3} Y_3} 
+ \frac{1}{\hat Z_{23} \alpha_{Y_2} Y_2 \tilde {\mathcal Y}_2} \Bigg] 
\ , 
\end{eqnarray}
where
\begin{eqnarray}
\hat Z_{23} &=& \frac{c_{12} + c_{1L} }{\alpha_{Y_1} Y_1 - \alpha_{Y_2} Y_2}   \ ,  \cr
\tilde {\mathcal Y}_{2} &=& c_{12} +c_{1L}+c_{2R}-  \hat Z_{23} \alpha_{Y_1} Y_1 + \hat Z_{23} \alpha_{Y_2} Y_2 - \alpha_{Y_3} Y_3  \ ,  \cr 
\tilde {\mathcal Y}_{3} &=& c_{12} + c_{1L} + c_{2L} -\hat Z_{23} \alpha_{Y_1} Y_1 \ . 
\end{eqnarray}

The fifth recursion term takes the form 
\begin{eqnarray}
\mathcal R_5^{{\text{double deformed loop}}} = \frac{\hat Z_{12}^2 }{\alpha_{ X_{1\bar 2}} X_{1\bar 2}}  \Bigg[\frac{1}{Z_{12}^2 \alpha_{Y_2} Y_2 \alpha_{Y_1} Y_1} 
+ \frac{1}{\tilde {\mathcal Y}_1 \tilde {\mathcal Y}_2} + \frac{1}{\hat Z_{12} \alpha_{Y_2} Y_2 \tilde {\mathcal Y}_2} + \frac{1}{\hat Z_{12} \alpha_{Y_1} Y_1 \tilde {\mathcal Y}_1} \Bigg] 
\ , 
\end{eqnarray}
where
\begin{eqnarray}
\hat Z_{12} &=& \frac{-c_{12} - c_{2R} +\alpha_{Y_3} Y_3}{\alpha_{Y_2} Y_2}   \ ,  \cr
\tilde {\mathcal Y}_{1} &=& c_{23} +c_{2L}+c_{3L}-  \hat Z_{12} \alpha_{Y_2} Y_2 \ ,  \cr  
\tilde {\mathcal Y}_{2} &=& c_{12} + c_{1L} + c_{2R} -\hat Z_{12} \alpha_{Y_1} Y_1 + \hat Z_{12} \alpha_{Y_2} Y_2 - \alpha_{Y_3} Y_3  \ . 
\end{eqnarray}

Finally, the sixth recursion term takes the form 
\begin{eqnarray}
\mathcal R_6^{{\text{double deformed loop}}} = \frac{\hat Z_{13}^2 }{\alpha_{ X_{13}} X_{13}}  \Bigg[\frac{1}{Z_{13} \alpha_{Y_1} Y_1 \alpha_{Y_3} Y_3} 
+ \frac{1}{\tilde {\mathcal Y}_1 \tilde {\mathcal Y}_3} + \frac{1}{\hat Z_{13} \alpha_{Y_1} Y_1 \tilde {\mathcal Y}_1} + \frac{1}{ \alpha_{Y_3} Y_3 \tilde {\mathcal Y}_3} \Bigg] 
\ , 
\end{eqnarray}
where
\begin{eqnarray}
\hat Z_{13} &=& \frac{c_{23} + c_{2L} +\alpha_{Y_3} Y_3}{\alpha_{Y_2} Y_2}   \ ,  \cr
\tilde {\mathcal Y}_{1} &=& c_{23} + c_{2L} + c_{3L} -\hat Z_{13} \alpha_{Y_2} Y_2  \ ,  \cr
\tilde {\mathcal Y}_{3} &=& c_{12} +c_{1L}+c_{2L}-  \hat Z_{13} \alpha_{Y_1} Y_1   \ . 
\end{eqnarray}

The explicit forms of all the six recursion terms even after  simplifications are lengthy and not very illuminating, so we collect them in the appendix \ref{4_loopExplicit}.

Summing these six recursion terms, we yield 
\begin{eqnarray}
\mathcal R^{{\text{deformed loop}}}  = \sum_{i=1}^{6} \mathcal R^{{\text{deformed loop deformed}}} _i  \ . 
\end{eqnarray}

It is very straightforward to see the expression $\mathcal R^{{\text{ deformed loop}}}$  is identical to stripped loop amplitude $\mathcal A_{{\text{deformed loop}}}$ \eqref{amp_4loop}.

\section{Interpretation of the recursion relation} \label{intep}
In the last section we worked out few examples of the recursion relation, while the more interesting question arise: How to interpret the different parts of the deformed amplitude formula \eqref{cfodass3} in the context of the recursion relation and geometric interpretation therein? We already shed some light to this in the previous section while discussing the recursion relation and few lower point examples.  In this section, we try to summarize it through a discussion of the recursion using the analytical structure of scattering amplitude.

\begin{itemize}
    \item Firstly, let us look at the denominator factor in the amplitude. The usual propagators $X_{ij}$
are present but in addition to that there are also product of $\alpha$’s in the denominator. We distribute them among the lower point functions when constructing the recursion relation.
It is very important to note that when we have more than one type of cubic coupling
terms present in the Lagrangian while constructing the higher point amplitudes, different
types of lower point amplitudes can be combined and that information is carried by the denominator factors.


 \item With a better understanding of the of the denominator factors; we turn our attention to the numerator factors of the amplitude. The natural thing is to try to absorb all the numerator factors into the lower point amplitudes which are the building blocks of the recursion framework, however it leads to inconsistencies (more on this later).Although it is already mentioned while deriving the recursion relation in Sec.3, that barring one $\alpha$ factor, all the remaining numerator factors can be absorbed in the factorized deformed amplitudes.
    As already pointed out for a theory with multiple scalar fields there are more than one type of three point building blocks which can be glued to construct the higher point amplitudes. 
    More precisely, due to the presence of more than one type of scalar fields, the BCFW-formula is also modified to be $\sum_i \mathcal A_{i,left}\times \mathcal A_{i,right}/X_i$ where $X_i$ are the various propagators (possibly representing different types of particles) that goes on-shell leading to factorization of the amplitude into the left and right sub-amplitudes. If one tries to distribute all the numerator factors among different building blocks, it leads to inconsistencies for the higher point amplitudes as we have already discussed for few examples in the last section.

    The above leads us to the conclusion that one of the numerator factor should be treated as an overall factor in the final deformed amplitude and not to be absorbed in the lower point amplitudes while factorizing.
    From the perspective of the Lagrangian of the field theory, one should consider it as taking a common factor out and rescaling the different coupling terms in the Lagrangian. One can always put back the factor in the final amplitude given the information of the initial triangulations involved to construct the associahedron.
     
     As already mentioned in sec.3, the numerator factors together acts as a scale factor to the original undeformed volume of the polytope when the deformed polytope is located in the same coordinate system  but the volume of the polytope remain unchanged in that transformed coordinate system \footnote{This is analogous to {\it active} and {\it passive} viewpoint of a vector transformation in a given coordinate system.}. We have showcased this fact in our literature through several lower point examples. 

    \item One question arise naturally that if we see from the Lagrangian perspective, the lowest point ({\it{i.e.}} three point in this case) amplitudes come with specific coupling constants. And as we construct the higher point amplitudes using these building blocks, the corresponding coupling constants $\lambda_i$ s are multiplied. The deformation parameters $\alpha$ s also do a similar job for our recursion relation to keep track of what kind of lower point functions are glued together to makeup the final amplitude. However one might ask why all the $\alpha$  factors cannot be absorbed into lower point amplitudes? To answer this pertaining question one should note the non-linear relation between the $\alpha$ s and the $\lambda_i$ s given in \eqref{coupling_rel}. If the relation between the deformation parameters and the coupling constants were linear in nature, one could have devised a one-to-one correspondence between the two and absorbed all the deformation parameters in the numerator inside the lower point functions while factorizing the amplitude.

\end{itemize}

\section{Geometric Picture : Projective Triangulations} \label{geo}

In this section we study the projective triangulation of the deformed associahedra and relate it to the recursion terms in section(\ref{recursion}). We show that different terms in the recursion relation correspond to projecting some of the facets of the associahedron to a lower dimensional geometry and triangulating the volume of the associahedron. To be more specific, each recursion terms is equal to the canonical function of the sub volume obtained by triangulating the associahedron. In other words, that is equal to the volume of the polytope dual to the projected sub volumes. Few explicit examples will be worked out to show this is the case. 

We note that the  canonical function of a simplex with vertices labelled by $Z_0,Z_1,\ldots,Z_m$ (in affine coordinates) is given by

\begin{eqnarray} \label{volume}
V \coloneqq[Z_0,Z_1,\ldots,Z_m]= \frac{\langle Z_0 Z_1 ... Z_{m}\rangle^m}{\prod_{i=0}^{m} \langle Y Z_1 \ldots \hat Z_i ... Z_{m}\rangle} \ . 
\end{eqnarray}

where angular brackets $\langle \ldots \rangle$ represent the determinant of the matrix formed using the rows $Z_i$, hat denotes the ommison of that particular vertex, and $Y=(1,X)$ with $X$ being the basis planar variables labelling the coordinates.
The case of $\phi^3$ theory with single field was given in \cite{Yang:2019esm}  where similar geometric picture of triangulating the associahedra was shown. In \cite{John:2020jww}, that picture was generalized to $\phi^p$ interactions with single scalar field.

\subsection{Four point amplitude in massless \texorpdfstring{$\phi^3$}
 ~ theory}

Let us consider the four-point amplitude. 
The deformed four-point amplitude in massless $\phi^3$ theory takes the form when we consider the initial traingulation to be along s-channel,
\begin{eqnarray}
\alpha  \Big(\frac{1}{\alpha s} + \frac{1}{\beta t}\Big) \ . 
\end{eqnarray}

If we project the 1-d associahedron along the s-axis, 
\begin{eqnarray}
Z_\star = \big(1, \frac{c}{\alpha}\big) \ ;  \  Z_1 = \big(1, 0\big) \ ;  \  Y = \big(1, s\big) \ . 
\end{eqnarray}

\begin{figure}[h] 
    \centering
\begin{tikzpicture}[scale=4, thick]
    \draw[->,black] (0,0) -- (1.3,0) node[below right, blue] {$s$};
    \draw[->,black] (0,0) -- (0,1.3) node[above left, blue] {$t$};

    \coordinate (O) at (0,0);
    \coordinate (A) at (1,0);      
    \coordinate (B) at (0,1);      

    \coordinate (C) at (0.8,0);    
    \coordinate (D) at (0,0.6);    

    \draw[thick, red] (A) -- (B); 
    \draw[thick, magenta] (C) -- (D); 
    \draw[ultra thick, magenta] (O) -- (C);
    \draw[thick, black] (O) -- (D);

    \node[below left, blue] at (O) {$(0,0)$};
    \node[below, blue] at (C) {$\left(\frac{c}{\alpha}, 0\right)$};
    \node[left, blue] at (D) {$\left(0, \frac{c}{\beta}\right)$};

    \node[rotate=-45, below right, red] at (0.50,0.53) {$s + t = c$};
    \node[rotate=-35, below right, magenta] at (0.1,0.5) {$\alpha s + \beta t = c$};
    \end{tikzpicture}
    
 \caption{The projectection of $1d$ associahedron on s-axis via recursion}
    \label{fig:enter-label}
\end{figure}

In our case $m=1$ and the deformed $1d$ associahedron is located in the s-t plane by the equation $\alpha s + \beta t =c$. 
Hence for our case the volume $V$ in \eqref{volume} takes the form

\begin{eqnarray} \label{v2}
V_t = \frac{\langle Z_\star Z_1 \rangle}{\langle Y Z_\star\rangle \langle Y Z_1 \rangle} = \alpha \Big( \frac{1}{\alpha s}+\frac{1}{\beta t} \Big) \ . 
\end{eqnarray}

So it seems we need to project along the proper direction to match the geometric picture with recursion term.

\subsection{Six point massless amplitude}
\label{geometry_6pt}

Here we look at the case of the deformed associahedron corresponding to six point amplitude. 
We consider the scaling of $X_{13}$ and $X_{14}$. We saw in section (\ref{six_point_recursion})  that the six point amplitude from the recursion relation has five terms and that their sum reproduces the correct deformed amplitude.

Let us now  look at the second term in the expression for the partial amplitude \eqref{second6point}. It takes the form :
\begin{equation}
\label{m49cube}
\mathcal R_2^d = \frac{c_{26}~ c_{36}~ \big(c_{35} - c_{36} + c_{46}\big)}{\alpha_{13}X_{13}~ \alpha_{36} X_{36}  ~\alpha_{46}X_{46}~ \mathcal A ~\mathcal B } \ , 
\end{equation}
where $\mathcal A := \big(c_{36}  \alpha_{13}X_{13} - c_{26}  \alpha_{14} X_{14}\big) $ and $ \mathcal B := \big(c_{36}-  \alpha_{15}X_{15} - c_{35}\big) $ are the spurious poles that appears in this term.
Note that all the poles are linear in the planar variables. We will now identify this term as the canonical function of the prism obtained by projecting the $X_{36}$ facet onto the $X_{13}X_{14}$ line. Following the discussion at the beginning of this section we identify the prism to be as in Figure \ref{fig:prismcube}.

\begin{figure}[htbp]
\centering
\begin{tikzpicture}[line join=round,line cap=round]

\draw[dashed,gray,thick]
(-1,5)--(8,6)--(8,10.2)--(-1,9.5)--cycle;
\draw[dashed,gray,thick]
(-1,5)--(1,1)--(10,2.3)--(8,6);
\draw[dashed,gray,thick]
(10,2.3)--(10.8,7.6)--(8,10.2);
\draw[thick] (0,5.1)--(1.7,2)--(7,2.8)--(8.3,5.45)--(8,6)--cycle;
\filldraw[fill=pink!30,draw=black,thick,opacity=0.6] (0,5.1)--(0,8)--(1.7,5.9)--(1.7,2); 
\filldraw[
  pattern=dots,
  pattern color=red] (0,5.1)--(0,8)--(1.7,5.9)--(1.7,2); 
\draw[thick] (1.7,5.9)--(3.5,7.5)--(7,7.8)--(7,2.8); 
\draw[thick] (0,8)--(3,9.5)--(4,9)--(3.5,7.5); 
\draw[thick] (3,9.5)--(8,9.8)--(8,6); 
\draw[newgreen,thick] (8,9.8)--(8,6); 
\draw[thick] (8,9.8)--(8.3,9.5)--(8.3,5.45); 
\draw[thick] (4,9)--(7.6,9.2)--(7,7.8); 
\draw[thick](7.6,9.2)--(8.3,9.5);
\begin{pgfonlayer}{background}
\fill[blue!20,opacity=0.7]
(1.7,2) -- (8,6) -- (0,5.1);
\fill[blue!20,opacity=0.3]
(1.7,2)--(1.7,5.9)--(8,8.2)--(8,6)--cycle;
\fill[blue!20,opacity=1.0]
(1.7,5.9)--(8,8.2)--(0,8)--cycle;
\fill[blue!20,opacity=0.6]
(0,8)--(0,5.1)--(8,6)--(8,8.2)--cycle;
\node at (4.8,5.9) {$X_{2,6}$};
\node at (8.2,3) {$X_{4,6}$};
\node at (-0.5,8.4) {$X_{1,3}$};
\node at (5.2,9.3) {$X_{1,5}$};
\node at (5.8,8.5) {$X_{2,5}$};
\node at (2.4,8) {$X_{3,5}$};
\node at (-0.1,4.5) {$X_{3,6}$};
\node at (7.6,7.2) {$X_{2,4}$};
 \node at (8.9,7.3) {$X_{1,4}$};

\end{pgfonlayer}

\end{tikzpicture}
\caption{\label{fig:6point_asso} Six point deformed associahedron and projection of $X_{36}$ facet onto $X_{13}X_{14}$ edge}
\end{figure}

\begin{figure}[htbp]
\centering
\begin{tikzpicture}
 \coordinate(Z1) at (0,0);
\node at (Z1) [left = 1mm of Z1] {$Z_1$};
\coordinate(Z2) at (5,0.7);
\node at (Z2) [right = .5mm of Z2] {$Z_2$};
\coordinate(Z*) at (1,-1.5);
\node at (Z*) [below = 1mm of Z*] {$Z_*$};
\coordinate(Z5) at (6.5,-.8);
\node at (Z5) [right = 1mm of Z5] {$Z_5$};
\coordinate(Z3) at (4.55,5.2);
\node at (Z3) [right= 1mm of Z3] {$Z_3$};
\coordinate(Z4) at (6.2,4.2);
\node at (Z4) [right=1mm of Z4] {$Z_4$};
\draw (Z1) -- (Z2);
\draw (Z*) -- (Z5);
\draw[newgreen]  (Z1) -- (Z*) node [midway,label = left:\textcolor{newgreen}{$X_{13},X_{14}$}] {};
\draw (Z2) -- (Z5);
\draw (Z2) -- (Z3);
\draw (Z4) -- (Z5);
\draw (Z4) -- (Z3);
\draw (Z4) -- (Z*);
\draw (Z1) -- (Z3);
\path[draw, fill=newgray, opacity =.15] (0,0) -- (5,0.7) -- (6.5,-.8) -- (1,-1.5) -- (0,0) -- cycle;
\node(X14) [inner sep = 0pt, newgray] at (3.3,-.4) {$X_{13}$};
\path[draw, fill=newred, opacity =.15] (5,0.7) -- (4.55,5.2) -- (6.2,4.2) -- (6.5,-.8) -- (5,0.7) -- cycle;
\node(X36) [inner sep = 0pt,newpurple] at (5.6,2.5) {$X_{36}$};
\path[draw, fill=newgreen, opacity =.15] (0,0) -- (5,0.7) -- (4.55,5.2) -- (0,0)-- cycle;
\node(X46) [inner sep = 0pt, newpink] at (3.2,1.8) {$X_{46}$};
\node(A) [inner sep = 0pt, label = left:{$\mathcal A = \big(c_{36}  \alpha_{13}X_{13} - c_{26}  \alpha_{14} X_{14}\big) $}] at (1.5, 4.5) {};
\draw[->] (4.2,3.3) to[bend right] (A);
\node(B) [inner sep = 0pt, label = right:{$\mathcal B := \big(c_{36} - c_{35} - \alpha_{15}X_{15}\big)  $}] at (6, -2) {};
\draw[->] (4.5,0) to[bend left] (B);
\end{tikzpicture}
\caption{\label{fig:prismcube}Prism formed by projecting $X_{36}$ facet onto $X_{13}X_{14}$ edge}
\end{figure}

We will now illustrate how we assign coordinates to the vertices of the prism. Let us choose the basis to be $\{X_{13},X_{14},X_{15}\}$.  Let us consider the vertex $\mathcal Z_3$. The coordinate for $\mathcal Z_{3} := \{X_{36},X_{46},\mathcal A\} $ is obtained by solving the following equations :
\begin{align*}
X_{36} &=  \frac{1}{\alpha_{36}} \big(-\alpha_{14} X_{14} + c_{36}\big)= 0 \ , \cr\\[5pt]
X_{46} &= \frac{1}{\alpha_{46}} \big(-\alpha_{15}X_{15} + c_{46}\big)=0 \ ,  \cr\\[5pt]
\mathcal A &= \big(c_{36}  \alpha_{13}X_{13} - c_{26}  \alpha_{14} X_{14}\big)=0 \ , \cr
\end{align*}
which gives  the coordinates of $\mathcal Z_3$ to be $(\frac{c_{26}}{\alpha_{13}},\frac{c_{36}}{\alpha_{14}},\frac{c_{46}}{\alpha_{15}}) $. The other coordinates $\mathcal Z _i$ can be found in a similar way by solving the appropriate linear equations. The affine coordinate $Z$ of the vertices of the prism is given by $Z := (1,\mathcal Z)$.
\\

We will now list  the coordinates of all the vertices of the prism :
 \begin{align}
 Z_{*}&\sim\{X_{13},X_{14},X_{15}\}=(1,0,0,\frac{c_{36}-c_{35}}{\alpha_{15}}) \ , \cr
Z_1&\sim\{X_{13},X_{14},X_{46}\}=(1,0,0,\frac{c_{46}}{\alpha_{15}}) \ , \cr
Z_2&\sim\{X_{13},X_{36},X_{46}\}=(1,0,\frac{c_{36}}{\alpha_{14}},\frac{c_{46}}{\alpha_{15}}) \ , \cr
Z_3&\sim\{X_{36},X_{46},\mathcal A\}=(1,\frac{c_{26}}{\alpha_{13}},\frac{c_{36}}{\alpha_{14}},\frac{c_{46}}{\alpha_{15}})\ , \cr
Z_4&\sim\{X_{36},\mathcal B, \mathcal A\}=(1,\frac{c_{26}}{\alpha_{13}},\frac{c_{36}}{\alpha_{14}},\frac{c_{36}-c_{35}}{\alpha_{15}}) \ , \cr
Z_5&\sim\{X_{13},X_{36},\mathcal B\}=(1,0,\frac{c_{36}}{\alpha_{14}},\frac{c_{36}-c_{35}}{\alpha_{15}}) \ . 
\end{align}
The canonical function of the prism is formed by adding the canonical function of the three simplices with vertices $\{Z_* Z_1 Z_2 Z_3\}$, $\{Z_* Z_2 Z_5 Z_4\}$ and $\{Z_* Z_2 Z_4 Z_3\}$  :
\begin{equation}
\underline{\Omega}_{36}=[Z_* Z_1 Z_2 Z_3]+[Z_* Z_2 Z_5 Z_4]+[Z_* Z_2 Z_4 Z_3] \ . 
\end{equation}
Using \eqref{volume} one can easily compute the deformed volume of the prism  $\underline{\Omega}_{36}n$.  And see that the result matches the second term in the partial stripped amplitude namely $\mathcal R_2^d $ as given in \eqref{m49cube} after the amplitude is multiplied by the overall volume factor $\underbrace{\alpha_{13}\alpha_{14}\alpha_{15}}$\footnote{Alternatively, one can rescale the coordinate axis $X_{ij}$ by $\alpha_{ij}$ and the volume $\Omega_{36}$ naturally accounts for the extra volume factor.}. 

 The pole $\mathcal A, \mathcal B$ correspond to the ``spurious boundary" within the associahedron introduced by triangulating it from inside via. the recursion. The salient feature of the recursion relation is that the residue at those spurious poles vanish when all the terms are added. This is extremely easy to visualize in the geometric picture we just mentioned. The spurious poles are nothing but internal surfaces introduced inside the associahedron to triangulate it and must cancel when all the triangulated volumes are added to recover the original associahedron.

 The other four terms in the recursion relation correspond to projecting $X_{26}, X_{25},X_{35}$ and $X_{24}$ faces onto the $X_{13}X_{14}$ line respectively. All of them contains the salient feature of having quadratic poles in the $X$ variables {\it c.f.} \eqref{6point_quadraticpole}. They correspond to curvy surfaces that arise while projecting onto the reference line. The reason being some of the edges of these faces are neither parallel nor perpendicular to the line onto which we are projecting. Moreover, the numerators also has $X$ dependent factors. Thus the geometries created by triangulating the associahedron in this specific way are although positive geometries but not simple polytopes. The appearance of ``curvy triangulation'' in the case of intermediate scalings of the basis variables were expected \cite{Yang:2019esm} and we indeed have them in the case of deformed associahedron.

\subsection{Three point loop amplitude}
We give a similar geometric interpretation of the recursion terms for the loop case. As we have already seen there are six terms in the recursion corresponding to $(Y_1,Y_2)$ rescaling. Among those we take the term coming from the dependent variable $X_{2\bar 3}$ for illustration \eqref{loop_term4}. 
\begin{equation}
\mathcal R_4^{{\text{double deformed loop}}} = \frac{c_{2L}c_{2R}~(c_{12}+c_{1L})}{\alpha_{Y_2}Y_2~\alpha_{Y_3}Y_3 ~\alpha_{X_{2\bar 3}}X_{2\bar 3}~\mathcal A ~\mathcal B} \ . 
\end{equation}

\begin{figure}[htbp]
\centering
\tdplotsetmaincoords{-70.5}{-32}
\begin{tikzpicture}[tdplot_main_coords,scale=2.]

\coordinate (Y1Y2Y3) at (0,0,0);
\coordinate (Y1Y2X12) at (0,0,2);
\coordinate (Y1Y3X13) at (0,2,0);
\coordinate (Y2Y3X23) at (2,0,0);
\coordinate (Y1Yt1X12) at (0,3,2);
\coordinate (Y1Yt1X13) at (0,3,1);
\coordinate (Y3Yt3X13) at (3,2,0);
\coordinate (Y3Yt3X23) at (3,1,0);
\coordinate (Y2Yt2X23) at (2,0,3);
\coordinate (Y2Yt2X12) at (1,0,3);
\coordinate (Yt1Yt2X12) at (1,3,3);
\coordinate (Yt1Yt3X13) at (3,3,1);
\coordinate (Yt2Yt3X23) at (3,1,3);
\coordinate (Yt1Yt2Yt3) at (3,3,3);

\draw (Y1Y2Y3) -- (Y1Y2X12) -- (Y1Yt1X12) --(Y1Yt1X13)--(Y1Y3X13) -- cycle;
\draw (Y1Y2Y3) -- (Y1Y2X12) -- (Y2Yt2X12) --(Y2Yt2X23)--(Y2Y3X23) -- cycle;
\draw (Y1Y2Y3) -- (Y2Y3X23) -- (Y3Yt3X23) --(Y3Yt3X13)--(Y1Y3X13) -- cycle;

\draw (Yt1Yt2Yt3) -- (Yt1Yt2X12) -- (Y1Yt1X12) --(Y1Yt1X13)--(Yt1Yt3X13) -- cycle;
\draw (Yt1Yt2Yt3) -- (Yt1Yt2X12) -- (Y2Yt2X12) --(Y2Yt2X23)--(Yt2Yt3X23) -- cycle;
\draw (Yt1Yt2Yt3) -- (Yt2Yt3X23) -- (Y3Yt3X23) --(Y3Yt3X13)--(Yt1Yt3X13) -- cycle;

\draw (Y1Y2X12) -- (Y1Yt1X12) -- (Yt1Yt2X12) --(Y2Yt2X12) -- cycle;
\draw (Y1Y3X13) -- (Y1Yt1X13) -- (Yt1Yt3X13) --(Y3Yt3X13) -- cycle;
\draw (Y2Y3X23) -- (Y2Yt2X23) -- (Yt2Yt3X23) --(Y3Yt3X23) -- cycle;

\draw[dashed,red,thick] (Y1Y2X12)--(0,0,3);
\draw[red,thick] (Y1Y2Y3)--(0,0,2);

\fill[blue!20,opacity=0.8]
(Y1Y2Y3) -- (Y2Y3X23) --(Y3Yt3X23) ;
\fill[blue!20,opacity=0.2]
(Y1Y2Y3) -- (0,0,3) --(Y2Yt2X23)--(Y2Y3X23)-- cycle ;
\fill[blue!20,opacity=0.8]
(0,0,3) -- (Y2Yt2X23) --(Yt2Yt3X23) ;
\fill[blue!20,opacity=0.4]
(Y1Y2Y3) -- (0,0,3) --(Yt2Yt3X23)--(Y3Yt3X23) -- cycle ;
\draw (0,1.6,1) node{$ Y_{1} $};
\draw (1,0,1.6) node{$ Y_{2} $};
\draw (1.6,1,0) node{$ Y_{3} $};

\draw (1.7,3,2) node{$ \tilde{Y}_{1} $};
\draw (2,1.4,3) node{$ \tilde{Y}_{2} $};
\draw (3,2,1.4) node{$ \tilde{Y}_{3} $};

\draw (0.5,1.5,2.5) node{$ X_{1 \bar 2}$};
\draw (1.5,2.5,0.5) node{$ X_{1 3}$};
\draw (2.5,0.5,1.5) node{$ X_{2 \bar 3}$};

\filldraw[
  pattern=dots,
  pattern color=black] (Y2Y3X23) -- (Y2Yt2X23) -- (Yt2Yt3X23) --(Y3Yt3X23) -- cycle;

\draw[fill opacity=0.25,fill=green](Y2Y3X23) -- (Y2Yt2X23) -- (Yt2Yt3X23) --(Y3Yt3X23) -- cycle;
\end{tikzpicture}
\caption{Geometric realisation of $D_3$ polytope and projection of $X_{2\bar 3}$ facet onto $Y_1Y_2$ edge }
\label{geometricy1y2y3}
\end{figure}

\begin{figure}[htbp]
\centering
\begin{tikzpicture}
 \coordinate(Z1) at (0,0);
\node at (Z1) [left = 1mm of Z1] {$Z_*$};
\coordinate(Z2) at (5,0.7);
\node at (Z2) [right = .5mm of Z2] {$Z_2$};
\coordinate(Z*) at (1,-1.5);
\node at (Z*) [below = 1mm of Z*] {$Z_1$};
\coordinate(Z5) at (6.5,-.8);
\node at (Z5) [right = 1mm of Z5] {$Z_5$};
\coordinate(Z3) at (4.55,5.2);
\node at (Z3) [right= 1mm of Z3] {$Z_3$};
\coordinate(Z4) at (6.2,4.2);
\node at (Z4) [right=1mm of Z4] {$Z_4$};
\draw (Z1) -- (Z2);
\draw (Z*) -- (Z5);
\draw[red]  (Z1) -- (Z*) node [midway,label = left:\textcolor{red}{$Y_{1},Y_{2}$}] {};
\draw (Z2) -- (Z5);
\draw (Z2) -- (Z3);
\draw (Z4) -- (Z5);
\draw (Z4) -- (Z3);
\draw (Z4) -- (Z*);
\draw (Z1) -- (Z3);
\path[draw, fill=newgray, opacity =.15] (0,0) -- (5,0.7) -- (6.5,-.8) -- (1,-1.5) -- (0,0) -- cycle;
\node(X14) [inner sep = 0pt, newgray] at (3.3,-.4) {$Y_{2}$};
\path[draw, fill=green, opacity =.15] (5,0.7) -- (4.55,5.2) -- (6.2,4.2) -- (6.5,-.8) -- (5,0.7) -- cycle;
\node(X23) [inner sep = 0pt,newpurple] at (5.6,2.5) {$ X_{2\bar3}$};
\path[draw, fill=newpurple, opacity =.15] (0,0) -- (5,0.7) -- (4.55,5.2) -- (0,0)-- cycle;
\node(Y3) [inner sep = 0pt, newpurple] at (3.2,1.8) {$Y_{3}$};
\node(A) [inner sep = 0pt, label = left:{$\mathcal A = \big(\alpha_{Y_2}(c_{12}+c_{1L}+c_{2L})Y_2-\alpha_{Y_1}c_{2L}Y_1\big) $}] at (1.5, 4.5) {};
\draw[->] (4.2,3.3) to[bend right] (A);
\node(B) [inner sep = 0pt, label = right:{$\mathcal B := \big(-c_{2R}+\alpha_{Y_3}Y_3\big)  $}] at (6, -2) {};
\draw[->] (4.5,0) to[bend left] (B);
\end{tikzpicture}
\caption{\label{fig:prismloop}Prism formed by projecting $X_{2\bar3}$ facet onto $Y_{1}Y_{2}$ edge}
\end{figure}

\FloatBarrier

The corresponding geometric projection is given in fig.\ref{fig:prismloop}\footnote{Note that the projection of the $X_{2\bar3}$ face onto the $Y_1Y_2$ line is a partial external triangulation of the $D_3$ polytope.}. It has two spurious boundaries corresponding to two spurious poles $(\mathcal A, \mathcal B)$. One can find out the coordinates of all the vertices of this prism and find the volume factor as in sec.\ref{geometry_6pt}. We will not be repeating the same as the procedure is identical in details. At the end the deformed canonical function of the prism exactly matches with the recursion term mentioned thus validating the expectation that the recursion terms corresponds to triangulation of the associahedron. Except the fourth term \eqref{loop_term4}, all the other five recursion terms contains spurious poles which are quadratic in the kinematic variables and thus triangulate the associahedron via curvy surfaces.

\section{EFT amplitudes}\label{EFT}
It was shown in \cite{Jagadale:2022rbl} that low energy effective field theory amplitude can be obtained by collapsing some of the propagators $\frac{1}{X_{ij}-m^2}$. This reduces some of the sub diagrams with cubic couplings to a lower point diagram with quartic interaction. 
If we start with two scalar fields $\phi_1$and $\phi_2$ with some interaction Lagrangian

\begin{equation}
    V(\phi_1,\phi_2)=\lambda_1 \phi_1^3+\lambda_2 \phi_1^2\phi_2 \ , 
\end{equation}

where $\phi_1$ is massless and $\phi_2$ is massive. Now one can take $(\lambda_2, m )\rightarrow \infty$ with $\frac{\lambda_2}{m}$ remaining finite and get the following effective potential,

\begin{equation}
    V_{\text{EFT}}(\phi_1)=\lambda_1\phi_1^3+g\phi_1^4 \ , 
\end{equation}

with $g=\frac{\lambda_2^2}{m^2}$ which remains finite as both $\lambda_1$ and $m$ taken to infinity.
From the perspective of Feynman diagrams, when we take $m \rightarrow \infty$, the massive propagators should collapse to give four point vertices of $\phi^4$ theory. Some of the $\phi^3$ diagrams might not contain the massive $\phi_2$ propagators and they remain untouched. Obviously, they don't contribute to the $\phi_4$ amplitudes.

This way of generating $\phi^p$ actions with $p > 3$ from a potential with more than one kind scalar fields (possibly massive), immediately raise the question whether there is a way to locate an accordiohedron (which is the positive geometry corresponding to $\phi^p$ interactions) starting from the $\phi^3$ associahedron. The answer to this question was attempted in \cite{Jagadale:2022rbl} and it was found that indeed one can locate the accordiohedron when looked at specific projection of the associahedron onto the region corresponding to $(\lambda_2, m) \rightarrow \infty$. Those geometries were also named as {\it projected accordiohedron}. 

In this paper, we ask the question how this picture of generating effective amplitudes translate to our recursion scheme. This is a key step in this program, the deformed realisation of the associahedron captures the amplitude of a field theory with multiple scalars interacting via cubic interactions. Then by taking appropriate limits in the parameters one can recover higher point interactions (even polynomial) effective amplitudes from it. Thus associahedron acts as as a universal polytope from which various amplitudes with theories with scalar particles can be obtained. Once we figure out how to implement the procedure to get EFT amplitude to the recursion scheme, that gives universality to the recursion method as well. The deformed recursion relation can work as a starting point for calculating any scalar amplitudes efficiently.  The expectation is to start with the recursion relation of the $\phi^3$ $n$-point deformed amplitude and when we fuse any one of the propagators it should lead to choosing only those recursion terms that contains that specific propagator ( all other terms should drop off as they do not contribute to EFT amplitude) and these terms also modify properly to give the appropriate $n$-point $\phi^4$ amplitude. The most economical way to achieve this is to take the residue of all the recursion terms at the pole given by the particular propagator going to zero!\\

Let us study one concrete example to see how this method works. Take the 5-point massive amplitude in $\phi^3$ theory as in appendix \ref{five_massive}. 
We consider $X_{13}$ to be the massive propagator {\it i.e.} $m_{13}^2=m^2$ and all the other propagation channel to be massless exchange. Under the rescaling $(X_{13}-m^2) \rightarrow z(X_{13}-m^2)$, we have two recursion terms. Now we want to obtain the effective 5-point amplitude with a $\phi^4$ verex and a $\phi^3$ vertex (of different strengths) when the massive $X_{13}$ propagator is fused by taking $m^2 \rightarrow \infty$. Only the first two terms in \eqref{5massive_amp} should contribute to the effective amplitude. And that is the exact contribution we get by taking the residue of the whole amplitude at $\underbrace{X_{13}-m^2}$ pole. Let us see how this reflects on the two recursion terms. 

\begin{align}
    \text{Recursion term 1},\mathcal {R}_1^{13}&= \frac{c_{25}(-c_{24}+c_{25}-c_{35})}{a_{13}\widetilde X_{13}~(c_{25}-a_{13}\widetilde X_{13})~(c_{25}-c_{24}-a_{14}X_{14})~(c_{35}-a_{14}X_{14})} \ , \cr
    \text{Recursion term 2},\mathcal {R}_2^{13} &= \frac{(c_{24} - c_{25}) (c_{24} + a_{14} X_{14})}{a_{13}\widetilde X_{13}~ a_{14} X_{14}~ (c_{24} - a_{13} \widetilde X_{13} + a_{14} X_{14})~ (c_{24} - c_{25} + a_{14} X_{14})} \ , 
\end{align}

where $\widetilde X_{13} =X_{13}-m^2$. Now we take the residue at $a_{13} \widetilde X_{13} \rightarrow 0$,

\begin{align}
    \text{Res}_{a13 \widetilde X_{13} \rightarrow 0} ~\mathcal {R}_1 &= \frac{(-c_{24}+c_{25}-c_{35})}{~(c_{25}-c_{24}-a_{14}X_{14})~(c_{35}-a_{14}X_{14})}\ , \cr
     \text{Res}_{a13 \widetilde X_{13} \rightarrow 0} ~\mathcal {R}_2 &=\frac{(c_{25} - c_{24}) }{ a_{14} X_{14}  (c_{25} - c_{24} - a_{14} X_{14})} \ . 
\end{align}

And once we add the two residues, we get the following amplitude,

\begin{eqnarray}
\label{phi4_4point}
\frac{c_{35}}{a_{14}X_{14}(c_{35}-a_{14}X_{14})} \xRightarrow{rewritten} \frac{1}{a_{14}X_{14}}+\frac{1}{a_{35}X_{35}} \ . 
\end{eqnarray}

This is exactly the $(\phi^3+\phi^4)$ five-point partial amplitude and corresponds to a accordiohedron with dissection $(14,35)$ which is the anticipated projected accordiohedron \cite{Jagadale:2021iab}.

One could instead choose to rescale $X_{14} \rightarrow z X_{14}$ and that would also give two recursion terms \eqref{5r0massive}, 

\begin{align}
     \text{Recursion term 1},\mathcal {R}_1^{14}&= \frac{(c_{25}-c_{24})}{a_{14} X_{14}~(c_{25}-a_{13}\widetilde X_{13})~(c_{24}-a_{13}\widetilde X_{13}+a_{14}X_{14})} \ , \cr
    \text{Recursion term 2},\mathcal {R}_2^{14} &= \frac{c_{25} c_{35}}{a_{13}\widetilde X_{13}~ a_{14} X_{14}~ (c_{25} - a_{13} \widetilde X_{13})~ (c_{35}- a_{14} X_{14})} \ . 
\end{align}
\\
Now, as the want to obtain the EFT amplitude, we take the residue at $a_{13} \widetilde X_{13} \rightarrow 0$ and immediately notice that $\mathcal R_1^{14}$ does not contribute as it has no pole at $a_{13} \widetilde X_{13} \rightarrow 0$ and only $\mathcal R_2^{14}$ contribute. That residue exactly contributes to the partial amplitude \eqref{phi4_4point}. So, the take home message here is that if we can choose the rescaling variable properly, when obtaining the effective amplitudes some of the recursion terms will not contribute at all and that makes the procedure calculationally more efficient. A wise choice of rescaling variables is such that the massive propagator appears in the minimum number of recursion terms. From the perspective of geometry, we should triangulate the associahedron in such a way that the facet corresponding to the massive pole is part of minimum number of triangulated cells. The geometric visualization of the associahedron can explicitly help in choosing the most efficient shift of the variables for the purpose of obtaining effective amplitudes without even looking at the explicit expression of the recursion terms.  

\section{Conclusions and outlook} \label{con}

In this literature we examined the applicability of the BCFW like recursion relations through a wide range of positive geometries. In particular, we studied the recursion relation adapted to deformed associahedron. 
Starting with the basic parts of ABHY associahedron for the case of scalar field theories, we discussed the structure of the poles that take place in the expression of the scattering amplitude and showed that how the associahedron acts as a positive geometry of a scalar field theory with cubic interactions. We confine our discussion only at the tree level planar amplitude. Afterwards, we introduced kinematic variables that have a one-to-one correspondence with the diagonals of $n$-gon which provides a clear picture about combinatorics of poles of the S-matrix.
Moreover, we reviewed how the canonical top form defined on the kinematic associahedron leads to the amplitude of planar $\phi^3$ theory. 

Later on we further clarify that an associahedron acts as a universal polytope and the scattering form on which contains the information about scalar theories with the most general kind of interaction.
We modified the kinematic variables appropriately to include the massive case as well. Then we do a rescaling of each of these variables due to which the hyperplane equations are modified. 
One can draw associahedron with new rescaled kinematic variables as the coordinates and will not see any difference compared to the old associahedron.
However if we consider the coordinate space of old kinematic variables, we obtain a deformed realisation of the assocaihedron.
From the top form defined upon the deformed associahedron one can extract the amplitude of a theory consisting of multiple scalar fields interacting via cubic vertices of different coupling strengths.  Then we discussed the non-linear relationship between coupling constants and the deformation parameter, which completes the review part.

Afterwards we focus on the recursion relations in the case of deformed amplitude through considering some 
specific examples.
Starting with the simplest case of four-point amplitude in massless $\phi^3$ theory, we further extend it to the six-point amplitude.
In the later case we uniformly rescaled two basis variables and discuss the recursion relation. 
Moreover we consider the loop amplitude in the deformed case.
For the case of loop amplitude we consider two loop variables and rescale them uniformly. 
In each case we found that in the expression of deformed amplitude along with the usual propagator factor there are also product of deformation factors present in the denominator. 
While constructing the recursion relation, we distribute them among the lower point functions in such a manner that it encode the information about the respective 
factorization channel. In addition, the deformed amplitude also contains an overall numerator factor consisting of product of deformation parameters, which in contrast to the denominator factors cannot be entirely distributed among the lower point functions. But if we extract one of those factors out in an appropriate way, the remaining factors can be distributed among the factorized lower point associahedrons in such a way that the canonical forms on those lower dimensional geometries indeed capture the deformed lower point amplitudes.

Towards the end, we discuss the corresponding geometric picture of the recursion terms as triangulation of the associahedron taking tree level four point and six point amplitudes , three point one-loop amplitude in massless $\phi^3$ theory as examples. The geometric picture is consistent with the recursion relation that we found earlier.

Finally  we discuss the case of obtaining effective field theory amplitudes from the Feynman diagrams by fusing one of the massive propagators to produce higher point vertices. In the geometric picture it translates to projecting the associahedron onto a specific surface given by that particular massive propagator going to zero. In this paper, we explore how to translate the idea of EFT amplitudes in the recursion scheme. We find the answer in terms of taking residue of the individual recursion terms at that particular propagator which is fused.
In particular, we illustrate the idea through the five point deformed amplitude in $\phi^3$ theory.

The positive geometry program makes us understand scattering amplitude as originating from geometrical objects rather than observable obtained from spacetime scattering processes. One may consider an abstract combinatorial object e.g. associahedron to be more fundamental. The basic properties of locality and unitarity of a S-matrix are encoded in the boundary structure of this geometrical object and the projective differential form defined on it. In this sense, one tells that rather being a fundamental entity, spacetime emerges from a much simpler and basic geometric objects. This approach brings about a radical conceptual change the way we used to conceive scattering amplitudes. But that is not the full story, moreover it also connects naturally to other approaches to obtain the amplitude. By extending the BCFW-like representation for a wider class of positive geometries, we also provide an efficient method to obtain the scattering amplitude of higher point associahedrons.

We end the section by pointing towards some of the future directions one can pursue. The most immediate one is to extend the recursion relation to the case of non-planar amplitudes. Recently, it was shown in \cite{Jagadale:2023hjr} that by summing over diffeomorphic copies of the associahedron exhaust all the pole structures of the $\phi^3$ theory including the non-planar ones. Since these copies are eventually associahedron combinatorially, the recursion should follow for them in a rather straightforward manner\footnote{It is worth mentioning that we have verified the recursion relation holds in its current form for the {\it associahedron block} or {\it colorful associahedron} which are also simple polytopes with similar combinatorial structures of an accordiohedron in general and can effectively capture scattering amplitudes involving exactly one massive propagator \cite{Jagadale:2021iab}.}. Extending the recursion scheme to gauge amplitudes has a roadblock due to the numerator factors present in it; recently in \cite{Laddha:2024qtn} the corolla polynomials were used to contract the canonical form on the associahedron to encode the tree level amplitude. The underlying geometry is still a simple polytope and a similar recursion scheme could be possible to engineer. The extension to gravity or string amplitudes could be interesting, albeit challenging. Recently a {\it curve integral} formula \cite{Arkani-Hamed:2023lbd,Arkani-Hamed:2023mvg} was suggested which capture the information about n-point scattering amplitude for arbitrary loop order and interestingly it does not refer to any Feynman diagram representation of the amplitude. The integral obtains the whole amplitude by separating it into a n-point tree level and a lower point loop amplitude, making it not only conceptually distinct but also calculationally tractable. It was already hinted that the curve integrals can be obtained recursively but adapting that to a BCFW like framework similar to the one discussed here is an important direction to pursue and we hope to report on this in the near future. A more ambitious connection to look at is the following: in general the relation between the couplings and deformation parameters is not straight
forward to obtain due to their non-linear nature. For an interaction Lagrangian of the form \eqref{int_lag}, in order to fix the various coupling constants, one needs the higher point deformed amplitude. This leaves us with a tempting idea if the recursion scheme which uses lower point amplitudes to build the higher point ones, can be faithful to get a precise relation (among coupling constants and deformation parameters) or at least put some constraints on its form.

\vskip .3in
 
\noindent {\bf\large Acknowledgement}

 \vskip .1in

\noindent We would like to thank Alok Laddha for discussing about the problem in an early stage of the project. We sincerely thank Dileep Jatkar and Anirban Basu for providing important insights about the results of the paper (especially about the interpretation of numerator factor) during a presentation of the preliminary findings. We are grateful to the anonymous referee who suggested many useful changes for the draft and helped us improve the manuscript in profound ways. 

\noindent The work of SRC is supported in part by the {\it{National Research Foundation of Korea}} (NRF) grant funded by the Korea government (MSIT) with grant number RS-2024-00449284, the Sogang University Research Grant of 202410008.01, the Basic Science Research Program of the {\it{National Research Foundation of Korea}} (NRF) funded by the Ministry of Education through the {\it{Center for Quantum Spacetime}} (CQUeST) with grant number RS-2020-NR049598.
SRC thank HRI for the visitor program during which parts of this work were carried out.

\appendix

\section{ An example: 5 point in massless \texorpdfstring{$\phi^3$}~ theory}
Let us now consider the tree level five point planar amplitude. We have the two basis variables $(X_{13},X_{14})$.
The convex realisation of the deformed associahedron is given by,
\begin{align}
    X_{35}&=\frac{1}{\alpha_{35}}(-\alpha_{14}X_{14}+c_{35}) \ , \cr 
    X_{25}&=\frac{1}{\alpha_{25}}(-\alpha_{13}X_{13}+c_{25}) \ , \cr
    X_{24}&=\frac{1}{\alpha_{24}}(\alpha_{14}X_{14}-\alpha_{13}X_{13}+c_{24}) \ . 
\end{align}
We do a single variable rescaling $X_{14} \rightarrow z X_{14}$. There are two terms in the recursion,
\begin{equation} \label{5r0}
    \mathcal A^{14}_5= \frac{z_{35}}{\alpha_{35} X_{35}}\left(\frac{1}{\alpha_{25}X_{25}}+\frac{1}{\alpha_{13}X_{13}}\right)+\frac{z_{24}}{\alpha_{24} X_{24}}\left(\frac{1}{\alpha_{14}X_{14}z_{24}}+\frac{1}{\alpha_{25}X_{25}}\right) \ , 
\end{equation}

where we have $z_{35}=\frac{c_{35}}{\alpha_{14}X_{14}}$ and $z_{24}=- \frac{c_{24}-\alpha_{13}X_{13}}{\alpha_{14}X_{14}}$.
\\

\begin{figure}[h]
\captionsetup{justification=centering}
\centering
\begin{minipage}{0.45\textwidth}
\centering
\begin{tikzpicture}
\coordinate(1) at (0,0) {};
\coordinate(2) at (0,4) {};
\coordinate(3) at (4,4) {};
\coordinate(4) at (4,1.2) {};
\coordinate(5) at (2.5,0) {};
\draw (1)--(2)  {};
\draw (2)--(3)  {};
\draw (3)--(4)  {};
\draw (4)--(5) {};
\draw (5)--(1) {};
\node[below left,newred] at (1) {$(0,0)$};
\node[below right,newred] at (5) {$(18,0)$};
\node[above left,newred] at (2) {$(0,50)$};
\node[above right,newred] at (3) {$(30,50)$};
\node[right,newred] at (4) {$(30,12)$};
\end{tikzpicture}
\caption*{\label{defass1}Deformed realization of the associahedron for $\alpha_{13}=\alpha_{14}=1$.}
\end{minipage}
\hspace{0.05\textwidth}
\begin{minipage}{0.45\textwidth}
\centering
\begin{tikzpicture}
\coordinate(1) at (0,0) {};
\coordinate(2) at (0,4) {};
\coordinate(3) at (4,4) {};
\coordinate(4) at (4,2) {};
\coordinate(5) at (2.5,0) {};
\draw (1)--(2)  {};
\draw (2)--(3)  {};
\draw (3)--(4)  {};
\draw (4)--(5) {};
\draw (5)--(1) {};
\node[below left,newred] at (1) {$(0,0)$};
\node[below right,newred] at (5) {$(6,0)$};
\node[above left,newred] at (2) {$(0,25)$};
\node[above right,newred] at (3) {$(10,25)$};
\node[right,newred] at (4) {$(10,6)$};
\end{tikzpicture}
\caption*{\label{defass2}Deformed realization of the associahedron for $\alpha_{13}=3,\alpha_{14}=2$.}
\end{minipage}
    \caption{Two distinct deformed realization of the associahedron in $(X_{13},X_{14})$ coordinate system with $c_{24}=18,c_{25}=30,c_{35}=50$.}
\label{5associa}
\end{figure}

In figure \ref{5associa}, we show two distinct realization of the 5-point associahedron with different values of $(\alpha_{13},\alpha_{14})$ . The values of $(\alpha_{24},\alpha_{25},\alpha_{35})$ do not show up in the deformed realization. As expected, the area of the first figure is 6 times of the second one and is exactly contributed to the ratio of volume factor $\underbrace{\alpha_{13}\alpha_{14}}$ in the two cases.

\subsection{Triangulation of the associahedron}

There are two recursion terms corresponding to dependent variables only $X_{35}$ and $X_{24}$. And geometrically that translates to outer triangulating the associahedron into a rectangle and a triangle by projecting $X_{35}$ and $X_{24}$ facets onto the $X_{14}$ line respectively Fig. \ref{prismloop}.

\begin{figure}[h]
\centering
\begin{tikzpicture}
\coordinate(1) at (0,0) {};
\coordinate(2) at (0,4) {};
\coordinate(3) at (4,4) {};
\coordinate(4) at (4,2) {};
\coordinate(5) at (2.5,0) {};
\draw (1)--(2) node[midway, label = left:\textcolor{newred}{$X_{13}$}] {};
\draw (2)--(3) node[midway, label = above:\textcolor{newred}{$X_{35}$}] {};
\draw (3)--(4) node[midway, label = right:\textcolor{newred}{$X_{25}$}] {};
\draw (4)--(5) node[midway, label = left:\textcolor{newred}{$X_{24}$}] {};
\draw (5)--(1) node[midway, label = below:\textcolor{newred}{$X_{14}$}] {};
\draw[dashed][newblue] (4,2) -- (4,0);
\draw[densely dotted][newblue] (2.5,0) -- (4,0);
\end{tikzpicture}
\caption{\label{prismloop}The edges $X_{35}$ and $X_{24}$ onto the $X_{14}$ edge via recursion.}
\end{figure}

First we consider the lower triangle with the coordinates as follows,

\begin{eqnarray}
Z_\star = \big(1, \frac{c_{24}}{\alpha_{13}}, 0\big) \ ;  \  Z_1 = \big(1, \frac{c_{25}}{\alpha_{13}}, 0\big) \ ;   \  Z_2 = \big(1, \frac{c_{25}}{\alpha_{13}}, \frac{c_{25} - c_{24}}{\alpha_{14}}\big) \ ; \  Y = \big(1, X_{13}, X_{14} \big) \ . 
\end{eqnarray}

Hence the volume will be

\begin{eqnarray} \label{vt}
V_T = \frac{\big(\langle Z_\star Z_1 Z_2 \rangle\big)^2}{\langle Y Z_\star Z_1\rangle \langle Y Z_1 Z_2 \rangle \langle Y Z_\star Z_2 \rangle} = \frac{\big(\alpha_{13} \alpha_{14} \big)\big(c_{25} - c_{24}\big)}{\alpha_{14} X_{14} \big(c_{24} - \alpha_{13} X_{13} + \alpha_{14} X_{14}\big) \big(c_{25} - \alpha_{13} X_{13}\big)} \ . 
\end{eqnarray}

Now for the rectangle part we break it into two triangle with $V_{T_1}$ and $V_{T_2}$.
Hence the volume of the rectangle is given by $V_R = V_{T_1}+  V_{T_2}$.

First we consider $V_{T_1}$.
In this case we have 

\begin{eqnarray}
Z_\star = \big(1, 0, 0\big) \ ;  \  Z_1 = \big(1, \frac{c_{25}}{\alpha_{13}}, 0\big) \ ;   \  Z_2 = \big(1, \frac{c_{25}}{\alpha_{13}}, \frac{c_{35}}{\alpha_{14}}\big) \ ; \  Y = \big(1, X_{13}, X_{14} \big) \ . 
\end{eqnarray}

Hence the volume will be

\begin{eqnarray} \label{vt1}
V_{T_1} = \frac{\big(\langle Z_\star Z_1 Z_2 \rangle\big)^2}{\langle Y Z_\star Z_1\rangle \langle Y Z_1 Z_2 \rangle \langle Y Z_\star Z_2 \rangle} 
= \frac{\big(\alpha_{13} \alpha_{14} \big)\big(c_{25} c_{35}\big)}{\alpha_{14} X_{14} \big(- \alpha_{13} X_{13} c_{35} + \alpha_{14} X_{14} c_{35} \big) \big(c_{25} - \alpha_{13} X_{13}\big)} \ . 
\end{eqnarray}

Now consider $V_{T_2}$.
In this case we have 

\begin{eqnarray}
Z_\star = \big(1, 0, 0\big) \ ;  \  Z_3 = \big(1, 0, \frac{c_{35}}{\alpha_{14}}\big) \ ;   \  Z_2 = \big(1, \frac{c_{25}}{\alpha_{13}}, \frac{c_{35}}{\alpha_{14}}\big) \ ; \  Y = \big(1, X_{13}, X_{14} \big) \ . 
\end{eqnarray}

Hence the volume will be

\begin{eqnarray} \label{vt2}
V_{T_2} = \frac{\big(\langle Z_\star Z_3 Z_2 \rangle\big)^2}{\langle Y Z_\star Z_3\rangle \langle Y Z_3 Z_2 \rangle \langle Y Z_\star Z_2 \rangle} =  \frac{\big(\alpha_{13} \alpha_{14} \big)\big(c_{25} c_{35}\big)}{\alpha_{13} X_{13} \big(- \alpha_{13} X_{13} c_{35} + \alpha_{14} X_{14} c_{25} \big) \big(-c_{35} + \alpha_{14} X_{14}\big)} \ .  
\end{eqnarray}

Therefore the volume of the Rectangle is given by

\begin{eqnarray} \label{vr}
V_R = V_{T_1} + V_{T_2} = -  \frac{c_{25} c_{35} \alpha_{13} \alpha_{14}}{\alpha_{13} X_{13} \alpha_{14} X_{14} \big(c_{25}-\alpha_{13}X_{13}\big) \big(c_{35}-\alpha_{14}X_{14}\big) } \ .
\end{eqnarray}

Hence finally the volume of our associahedron will be given by 

\begin{eqnarray} \label{v}
V &=& V_R  \ -  \ V_T    \cr
 &=&-\frac{ \alpha_{13} \alpha_{14}}{\alpha_{14} X_{14} \big(c_{25}-\alpha_{13}X_{13}\big) } \Bigg[  \frac{c_{25} c_{35}}{\alpha_{13} X_{13}  \big(c_{35}-\alpha_{14}X_{14}\big) } + \frac{c_{25} - c_{24}}{c_{24} - \alpha_{13} X_{13} + \alpha_{14}X_{14} } \Bigg]  \ . 
\end{eqnarray}

The above expression matches with the expression that we get from the recursion relation as given in \eqref{5r0}.

\section{Five point amplitude in massive \texorpdfstring{$\phi^3$}~ theory}
\label{five_massive}
 We also note down the recursion for the five point deformed amplitude in the case of massive $\phi^3$ theory. This will be useful for the discussion of EFT amplitude in sec.\ref{EFT}.

For the five point amplitude in massive $\phi^3$ theory we have 

\begin{eqnarray}
X_{35} &=& \frac{1}{\alpha_{35}} \Big(-\alpha_{14} \big(X_{14} - m_{14}^2\big)+ c_{35}\Big) + m_{35}^2 \ ,  \cr
&&\cr
X_{25} &=& \frac{1}{\alpha_{25}} \Big(-\alpha_{13} \big(X_{13} - m_{13}^2\big)+ c_{25}\Big) + m_{25}^2  \ ,  \cr
&&\cr
X_{24} &=& \frac{1}{\alpha_{24}} \Big(\alpha_{14} \big(X_{14} - m_{14}^2\big) - \alpha_{13} \big(X_{13} - m_{13}^2\big) + c_{24} \Big) + m_{24}^2  \ , 
\end{eqnarray}

Under the rescaling $X_{14} \rightarrow zX_{14}$,  the recursion relation takes the form 
{\scriptsize
\begin{equation} \label{5r0massive}
    \mathcal A^{14}_5= \frac{z_{35m}}{\alpha_{35} (X_{35}-m_{35}^2)}\left(\frac{1}{\alpha_{25}(X_{25}-m_{25}^2)}+\frac{1}{\alpha_{13}(X_{13}-m_{13}^2)}\right)+\frac{z_{24m}}{\alpha_{24} (X_{24}-m_{24}^2)}\left(\frac{1}{\alpha_{14}(X_{14}-m_{14}^2)z_{24}}+\frac{1}{\alpha_{25}(X_{25}-m_{25}^2)}\right)  
\end{equation}
}
with  $z_{35 m} = \frac{c_{35}}{\alpha_{14} \big(X_{14}-m_{14}^2\big)}$ and $z_{24 m} = \frac{-c_{24}+\alpha_{13}\big(X_{13}-m_{13}^2\big)}{\alpha_{14}\big(X_{14}-m_{14}^2\big)}$.

\quad

After simplification the five-point massive stripped amplitude becomes

\begin{eqnarray}
\label{5massive_amp}
\mathcal A_5 &=& \frac{1}{\alpha_{13} \big(X_{13} - m_{13}^2\big) \alpha_{14} \big(X_{14} - m_{14}^2\big)} + \frac{1}{\alpha_{13} \big(X_{13} - m_{13}^2\big) \alpha_{35} \big(X_{35} - m_{35}^2\big)} \cr
&& \cr
&&+  \ \frac{1}{\alpha_{25} \big(X_{25} - m_{25}^2\big) \alpha_{35} \big(X_{35} - m_{35}^2\big)} +  \frac{1}{\alpha_{25} \big(X_{25} - m_{25}^2\big) \alpha_{24} \big(X_{24} - m_{24}^2\big)} \cr
&& \cr
&&+ \  \frac{1}{\alpha_{24} \big(X_{24} - m_{24}^2\big) \alpha_{14} \big(X_{14} - m_{14}^2\big)}  \ . 
\end{eqnarray}

\section{Volume of deformed lower point associahedrons}\label{volume_cal}

In this section,we explicitly show for two cases how the canonical form of the facet of a three dimensional deformed associahedon is given by product of canonical forms of two lower point deformed associahedrons.  

First consider taking the $X_{36}$ propagator onshell, that amounts to landing on the $X_{36}$ surface of the associahedron in Fig.\ref{fig:6point_asso}. Now taking $X_{36} \to 0$ in the amplitude, breaks the six point amplitude to two four point amplitudes. Correspondingly, the concerned facet is a quadrilateral and given by product of two line segments which are four point associahedrons. We support our claim by explicitly calculating the canonical form on this surface and then showing that form indeed factorize into two lower forms. Firstly, we note down the vertices of the $X_{36}$ face in the $(X_{13},X_{15})$ coordinate system (for this face $X_{14} = \frac{c_{36}}{\alpha_{14}}$ for all the verices, so we omit it),

\begin{equation}
    z_1 : \left(0,\frac{c_{46}}{\alpha_{15}}\right)~~ z_2 : \left(\frac{c_{26}}{\alpha_{13}},\frac{c_{46}}{\alpha_{15}}\right)~~ z_3 : \left(\frac{c_{26}}{\alpha_{13}},\frac{c_{36}-c_{35}}{\alpha_{15}}\right)~~ z_4 : \left(0,\frac{c_{36}-c_{35}}{\alpha_{15}}\right)
\end{equation}

By making the coordinates projective {\it i.e.} $(1,z)$ coordinates, we can calculate the canonical function on this surface by diving it into two triangles,

\begin{equation}
    \underline{\Omega}(X_{36})= [Z_1Z_3Z_4]+[Z_1Z_2Z_3]
\end{equation}

One can do the calculation using the \eqref{volume}, we just quote the final answer,
\begin{equation}
     \underline{\Omega}(X_{36})=\frac{\alpha_{13}\alpha_{15}c_{26}(c_{36}-c_{35}-c_{46})}{\alpha_{13}X_{13}~\alpha_{26}X_{26}~\alpha_{46}X_{46}~\alpha_{35}X_{35}}
\end{equation}

We can write down the corresponding canonical form by multiplying the canonical function with the volume form,

\begin{align}
    \Omega(X_{36})&= \frac{\alpha_{13}\alpha_{15}c_{26}(c_{35}-c_{36}+c_{46})}{\alpha_{13}X_{13}~\alpha_{26}X_{26}~\alpha_{46}X_{46}~\alpha_{35}X_{35}} dX_{13} \wedge dX_{15} \cr
    &\cr
    &= \left(\frac{1}{\alpha_{13}X_{13}}+\frac{1}{\alpha_{26}X_{26}}\right) \alpha_{13}dX_{13} ~\times~ \left(\frac{1}{\alpha_{35}X_{35}}+\frac{1}{\alpha_{46}X_{46}}\right) \alpha_{35}dX_{35} 
\end{align}

On the second equality we have written the canonical form as product of two canonical forms which represent two one-dimensional associahedrons corresponding to four point amplitudes in which the six point amplitude factorizes when we take the $X_{36}$ propagator onshell (see Fig.\ref{x36}). Also note that we have used the fact on this surface  $\alpha_{15}dX_{15}=\alpha_{35}dX_{35}$. To complete the discussion, we know the six point deformed amplitude contains a product of $\underbrace{\alpha_{13}\alpha_{14}\alpha_{15}}$ in the numerator but the factorized associahedron carries only the $\alpha_{13}$ and $\alpha_{15}$ factors, so $\alpha_{14}$ is the remaining factor that is needed to be entered as an overall factor (see \eqref{factor_deform}) to construct the six point amplitude.

\begin{figure}[htbp]
  \centering
\begin{tikzpicture}[scale=0.8, every node/.style={font=\small}]

\coordinate (A) at (0,0);
\coordinate (B) at (2,0);
\coordinate (C) at (4,0);
\coordinate (D) at (6,0);

\draw (A)--(B)--(C)--(D);

\draw (A)--(-1,0.7) node[left] {4};
\draw (A)--(-1,-0.7) node[left] {3};

\draw (D)--(7,0.7) node[right] {1};
\draw (D)--(7,-0.7) node[right] {2};

\draw (B)--(2,1) node[above] {5};
\draw (C)--(4,1) node[above] {6};

\node at (1,-0.4) {$X_{35}$};
\node at (3,-0.4) {$X_{36}$};
\node at (5,-0.4) {$X_{13}$};

\draw[->, line width=1.2pt] (7.4,0) -- (9.2,0);
\node[green!60!black, above] at (8.3,0) {$X_{36}\to 0$};

\coordinate (E) at (10.8,0);
\coordinate (F) at (12.8,0);

\draw (E)--(F);

\draw (E)--(9.8,0.7) node[left] {4};
\draw (E)--(9.8,-0.7) node[left] {3};

\draw (F)--(13.8,0.7) node[right] {5};
\draw (F)--(13.8,-0.7) node[right] {I};

\node at (11.8,-0.4) {$X_{35}$};

\node at (14.6,0) {$\times$};

\coordinate (G) at (16.2,0);
\coordinate (H) at (18.2,0);

\draw (G)--(H);

\draw (G)--(15.2,0.7) node[left] {6};
\draw (G)--(15.2,-0.7) node[left] {I};

\draw (H)--(19.2,0.7) node[right] {1};
\draw (H)--(19.2,-0.7) node[right] {2};

\node at (17.2,-0.4) {$X_{13}$};

\begin{scope}[yshift=-2.2cm, xshift=10.8cm]
\draw[red, line width=1.2pt] (0,0)--(3,0);
\node at (0,-0.4) {$X_{35}$};
\node at (3,-0.4) {$X_{46}$};

\draw[red, line width=1.2pt] (4.8,0)--(7.8,0);
\node at (4.8,-0.4) {$X_{13}$};
\node at (7.8,-0.4) {$X_{26}$};
\end{scope}

\end{tikzpicture}
\caption{\label{x36} Factorization of a  six point diagram in two four point diagrams in the $X_{36} \to 0 $ channel and the corresponding one dimensional associahedrons are drawn in red.}
\end{figure}

Next we consider the $X_{25}$ face, which is again a product of two one dimensional associahedrons. But unlike the $X_{36}$ surface, it is not perpendicular to any of the coordinate axes. As a consequence, at first it seems the canonical form on this face contains wedge product of all the $(dX_{13},dX_{14},dX_{15})$. But on the $X_{25}$ surface we also have $\alpha_{13}~dX_{13}=\alpha_{15}~dX_{15}$. So, we can eliminate either of $X_{13}$ and $X_{15}$. Using the same procedure as the above example one can calculate the canonical form, we quote the final expression,

\begin{align}
    \Omega(X_{25})&= \frac{\alpha_{13}\alpha_{14}(c_{25}-c_{26})(c_{35}-c_{25}+c_{24})}{\alpha_{15}X_{15}~\alpha_{26}X_{26}~\alpha_{24}X_{24}~\alpha_{35}X_{35}} dX_{13} \wedge dX_{14} \cr
    &\cr
    &= \left(\frac{1}{\alpha_{26}X_{26}}+\frac{1}{\alpha_{15}X_{15}}\right) \alpha_{26}dX_{26} ~\times~ \left(\frac{1}{\alpha_{24}X_{24}}+\frac{1}{\alpha_{35}X_{35}}\right) \alpha_{24}dX_{24} 
\end{align}

\begin{figure}[htbp]
  \centering
\begin{tikzpicture}[scale=0.8, every node/.style={font=\small}]

\coordinate (A) at (0,0);
\coordinate (B) at (2,0);
\coordinate (C) at (4,0);
\coordinate (D) at (6,0);

\draw (A)--(B)--(C)--(D);

\draw (A)--(-1,0.7) node[left] {3};
\draw (A)--(-1,-0.7) node[left] {2};

\draw (D)--(7,0.7) node[right] {6};
\draw (D)--(7,-0.7) node[right] {1};

\draw (B)--(2,1) node[above] {4};
\draw (C)--(4,1) node[above] {5};

\node at (1,-0.4) {$X_{24}$};
\node at (3,-0.4) {$X_{25}$};
\node at (5,-0.4) {$X_{26}$};

\draw[->, line width=1.2pt] (7.4,0) -- (9.2,0);
\node[green!60!black, above] at (8.3,0) {$X_{25}\to 0$};

\coordinate (E) at (10.8,0);
\coordinate (F) at (12.8,0);

\draw (E)--(F);

\draw (E)--(9.8,0.7) node[left] {3};
\draw (E)--(9.8,-0.7) node[left] {2};

\draw (F)--(13.8,0.7) node[right] {4};
\draw (F)--(13.8,-0.7) node[right] {I};

\node at (11.8,-0.4) {$X_{24}$};

\node at (14.6,0) {$\times$};

\coordinate (G) at (16.2,0);
\coordinate (H) at (18.2,0);

\draw (G)--(H);

\draw (G)--(15.2,0.7) node[left] {5};
\draw (G)--(15.2,-0.7) node[left] {I};

\draw (H)--(19.2,0.7) node[right] {6};
\draw (H)--(19.2,-0.7) node[right] {1};

\node at (17.2,-0.4) {$X_{26}$};

\begin{scope}[yshift=-2.2cm, xshift=10.8cm]
\draw[red, line width=1.2pt] (0,0)--(3,0);
\node at (0,-0.4) {$X_{24}$};
\node at (3,-0.4) {$X_{35}$};

\draw[red, line width=1.2pt] (4.8,0)--(7.8,0);
\node at (4.8,-0.4) {$X_{26}$};
\node at (7.8,-0.4) {$X_{15}$};
\end{scope}

\end{tikzpicture}
\caption{\label{x25} Factorization of a  six point diagram into two four point diagrams in the $X_{25} \to 0 $ channel and the corresponding one dimensional associahedrons are drawn in red.}
\end{figure}

The diagramatics is given in Fig.\ref{x25}. In this case, we need to add $\alpha_{15}$ as an overall factor to recover the six point amplitude via. recursion formula. One can run a similar exercise for the pentagonal faces but in that case the calculation of the canonical form is simpler as the factorized amplitude correspond to a product of two dimensional associahedron (pentagon) and a zero dimensional associahedron (point).

\section{Explicit forms of the recursion terms in 3 point loop amplitude}
\label{4_loopExplicit}

{\scriptsize
\begin{eqnarray}
\mathcal R_1^{{\text{double deformed loop}}} &=& \Bigg[\big(c_{23} + c_{2L} +c_{3L}\big) \big(-\alpha_{Y_2} \big(c_{12} + c_{1L} + c_{2L}\big) \big(c_{12} + c_{23} + c_{2L} + c_{2R}\big) Y_2 \cr
&& \big(c_{12} + c_{23} + c_{2L} +c_{2R} +c_{3L} +c_{1L} - \alpha_{Y_3} Y_3\big) +  \alpha_{Y_1} Y_1 \big(c_{23} +c_{2L}+c_{3L}\big) \cr
&&\big(c_{12}^2 + c_{23} c_{2L} + c_{2L}^2 + c_{2L} c_{2R} 
+c_{2L} c_{3L} -c_{3L}c_{1L} - \alpha_{Y_3} c_{2L} Y_3 + \alpha_{Y_3} c_{1L} Y_3 \cr
&&+ c_{1L} \big(c_{23} + c_{2L} + c_{2R} + c_{3L} -  \alpha_{Y_3} Y_3\big) + c_{12} \big(c_{1L}+c_{23} +2c_{2L} +c_{2R} \cr
&&+c_{3L} -  \alpha_{Y_3} Y_3\big)\big)\big)\Bigg]  
\Bigg[ \alpha_{Y_1}Y_1~ \alpha_{\tilde Y_1}  \tilde Y_1 \big(\alpha_{Y_1} \big(c_{23} +c_{2L} +c_{3L}\big) Y_1 - \alpha_{Y_2} \big(c_{12} \cr
&&+c_{1L} +c_{2L} \big) Y_2\big)  \big(c_{3L} -  \alpha_{Y_3} Y_3\big) \big(c_{12} +c_{23} +c_{2L} + c_{2R} + c_{3L} - \alpha_{Y_3} Y_3\big) \cr
&&\big(\alpha_{Y_1} \big(c_{23} +c_{2L} + c_{3L}\big) Y_1 - \alpha_{Y_2} Y_2 \big(c_{12} +c_{23} + c_{2L} + c_{2R} + c_{3L} + c_{1L} \cr
&& -  \alpha_{Y_3} Y_3\big)\big)\Bigg]^{-1} \ . 
\end{eqnarray}
}

{\scriptsize
\begin{eqnarray}
\mathcal R_2^{{\text{double deformed loop}}} &=& \Bigg[\big(c_{12} + c_{1L} + c_{2L}\big)^2 \big(-\alpha_{Y_2}^2 c_{23} \big(c_{12} + c_{1L} + c_{2L}\big)^2 Y_2^2 + \alpha_{Y_1} \alpha_{Y_2} \cr
&&c_{23} \big(c_{12} + c_{1L} +c_{2L} \big)Y_1 Y_2 \big(c_{23} + 2c_{2L}  - c_{2R} +c_{3L} + \alpha_{Y_3} Y_3\big)\cr
&&+ \alpha_{Y_1}^2 Y_1^2 \big(c_{23}^2 \big(-c_{2L} +c_{2R}\big)+\alpha_{Y_3}Y_3 c_{2R}c_{3L} - c_{23}\big(c_{2L} -c_{2R}\big) \big(c_{2L} + c_{3L} + \alpha_{Y_3} Y_3\big)\big)\big)\Bigg] \cr
&&\Bigg[\alpha_{Y_3} Y_3 ~ \alpha_{\tilde Y_3}\tilde Y_3 \big(-\alpha_{Y_1} c_{2L} Y_1 + \alpha_{Y_2} \big(c_{12} + c_{1L} + c_{2L}\big)Y_2\big) \cr
&&\big(-\alpha_{Y_1} \big(c_{23} +c_{2L}+c_{3L}\big)Y_1 + \alpha_{Y_2} \big(c_{12} + c_{1L} + c_{2L} \big) Y_2\big)  \big(-\alpha_{Y_2} \big(c_{12} + c_{1L} + c_{2L}\big)\cr
&&Y_2 + \alpha_{Y_1}Y_1 \big(c_{23} + c_{2L}-\alpha_{Y_3} Y_3\big)\big) \big) \Bigg]^{-1} 
\end{eqnarray}
}

{\scriptsize
\begin{eqnarray}
\mathcal R_3^{{\text{double deformed loop}}} &=& \Bigg[\big(c_{12}+c_{1L}-\alpha_{Y_3}Y_3\big)\big(\alpha_{Y_1}^2 Y_1^2 c_{12} \big(c_{23} + c_{12}+c_{3L}\big) \big(c_{2L}-c_{2R}+\alpha_{Y_3}Y_3\big) + \alpha_{Y_2}^2 Y_2^2 \big(c_{12}c_{1L}c_{2L}+c_{12} \cr
&&\big(c_{23}+c_{2L}+c_{3L}\big) \big(c_{2R}-\alpha_{Y_3}Y_3\big)+c_{2L} \big(c_{1L}+c_{23}+c_{2L}+c_{3L}\big)\big(c_{1L}+c_{2R}-\alpha_{Y_3}Y_3\big)\big)+\alpha_{Y_1}Y_1 \alpha_{Y_2}Y_2\cr
&&\big(c_{12}^2 c_{2L}-c_{2L} \big(c_{1L}+c_{2R}-\alpha_{Y_3}Y_3\big)\big(c_{23}+c_{2L}-c_{2R}+c_{3L}-\alpha_{Y_3}Y_3\big)+c_{12} \big(c_{1L}c_{2L} + c_{2L}^2+c_{2L} c_{3L}-\cr
&&-2c_{2R}c_{3L}+2\alpha_{Y_3}Y_3 c_{3L} +c_{23} \big(c_{2L}-2c_{2R}+2\alpha_{Y_3}Y_3\big)\big)\big)\big)\Bigg]
\Bigg[\alpha_{Y_2}Y_2~ \alpha_{\tilde Y_2} \tilde Y_2  \big(c_{2R}-\alpha_{Y_3}Y_3\big) \big(\alpha_{Y_2}Y_2 c_{1L} \cr
&&+\alpha_{Y_1}Y_1 \big(c_{12}+c_{2R}-\alpha_{Y_3}Y_3\big)\big) \big(-\alpha_{Y_1} Y_1 \big(c_{23}+c_{2L}+c_{3L}\big)
+\alpha_{Y_2} Y_2\big(c_{12}+c_{1L}+c_{23}+c_{2L}+c_{2R} \cr
&&+c_{3L}-\alpha_{Y_3}Y_3\big)\big)\Bigg]^{-1}
\end{eqnarray}
}
{\small
\begin{equation}
\label{loop_term4}
\mathcal R_4^{{\text{double deformed loop}}} = \frac{c_{2L}c_{2R}(c_{12}+c_{1L})}{\alpha_{Y_2}Y_2~\alpha_{Y_3} Y_3 ~\alpha_{X_{2\bar 3}}X_{2\bar 3}~(-c_{2R}+\alpha_{Y_3}Y_3)~\big(-c_{2L}\alpha_{Y_1}Y_1+(c_{12}+c_{1L}+c_{2L})\alpha_{Y_2}Y_2\big)}
\end{equation}
}
{\small
\begin{eqnarray}
\mathcal R_5^{{\text{double deformed loop}}} &=& \bigg[c_{1L}(c_{23}+c_{2L}+c_{3L})\bigg] \bigg[\alpha_{Y_1} Y_1 (c_{12}+c_{23}+c_{2L}+c_{2R}+c_{3L}-\alpha_{Y_3} Y_3) \cr
&& \alpha_{X_{1\bar 2}}X_{1\bar 2} (\alpha_{Y_1} Y_1 c_{12}+ \alpha_{Y_1} Y_1 c_{2R}+\alpha_{Y_2} Y_2 c_{1L} \cr
&&-\alpha_{Y_1} Y_1 \alpha_{Y_3} Y_3)\bigg]^{-1} \ . 
\end{eqnarray}
}

{\small
\begin{eqnarray}
\mathcal R_6^{{\text{double deformed loop}}} &=& \bigg[(c_{12}+c_{1L} +c_{2L} ) c_{3L} (c_{23}+c_{2L} +\alpha_{Y_3} Y_3) \bigg] \bigg[\alpha_{Y_1} Y_1 \alpha_{Y_3} Y_3 (-c_{3L}+\alpha_{Y_3} Y_3) \cr
&& \alpha_{X_{13}} X_{13} Y_1 (-\alpha_{Y_2} Y_2(c_{12}+c_{1L}+c_{2L})Y_2 + \alpha_{Y_1} Y_1 (c_{23} \cr
&&+ c_{2L}+ \alpha_{Y_3} Y_3))\bigg]^{-1} \ . 
\end{eqnarray}
}

\end{document}